\pdfoutput=1
\documentclass[aip,pop,amsmath,reprint]{revtex4-1}
\usepackage{natbib}
\usepackage{hyperref}
\usepackage{txfonts}
\usepackage{graphicx}
\usepackage{bm}
\usepackage{ctable}
\usepackage{url}
\usepackage{etoolbox}
\usepackage{cleveref}
\crefname{section}{\S\!}{\S\S\!}
\Crefname{section}{Section}{Sections}
\crefname{appendix}{App.}{Apps.}
\Crefname{appendix}{Appendix}{Appendices}
\crefname{equation}{Eq.}{Eqs.}
\crefname{subequation}{Eqs.}{Eqs.}
\Crefname{equation}{Equation}{Equations}
\crefname{figure}{Fig.}{Figs.}
\Crefname{figure}{Figure}{Figures}
\Crefname{table}{Table}{Tables}

\newcommand{\athenapp}{\texttt{Athena++}} 
\newcommand{\companion}{S+22} 
\newcommand{\mallet}{M+21} 
\newcommand{\bolddot}{\bm{\cdot}} 
\newcommand{\boldtimes}{\bm{\times}} 
\newcommand{\parfrac}[2]{\frac{\partial #1}{\partial #2}}  
\newcommand{\B}{\bm{B}} 
\newcommand{\mB}{\overline{\bm{B}}} 
\newcommand{\vel}{\bm{u}} 
\newcommand{\va}{\bm{v}_{\rm A}} 
\newcommand{\vamag}{v_{\rm A}} 
\newcommand{\bz}{\bm{z}}  
\newcommand{\ph}{\hat{\bm{p}}} 
\newcommand{\AdB}{A_{\delta\B}}
\newcommand{\Adu}{A_{\delta\vel}}
\newcommand{\tildenabla}{\widetilde{\nabla}} 
\newcommand{\parkerangle}{\Phi_{\text{P}}} 
\newcommand{\CB}{C_{\B^2}} 
\newcommand{\clockangle}{\vartheta_{\rm CA}} 
\newcommand{\deflectangle}{\vartheta_z} 
\newcommand{\bsf}[1]{\mathsf{\bm{#1}}}  
\newcommand{\BLambda}{\bm{\Lambda}} 
\newcommand{\chg}{\tilde{\bz}^+_{\rm WA}} 
\newcommand{\chf}{\tilde{\bz}^-_{\rm WA}}
\newcommand{\chgf}{\tilde{\bz}^\pm_{\rm WA}}
\newcommand{\defaulthires}{\textit{CubicHR}}
\newcommand{\parkerhires}{\textit{CubicParker15HR}}
\newcommand{\betalow}{\textit{Cubic-$\beta_0$0.16}}
\newcommand{\betahigh}{\textit{Cubic-$\beta_0$1}}
\newcommand{\chilow}{\textit{Cubic-$\chi_0$0.2}}
\newcommand{\chihigh}{\textit{Cubic-$\chi_0$1.4}}

\makeatletter
\def\@email#1#2{%
 \endgroup
 \patchcmd{\titleblock@produce}
  {\frontmatter@RRAPformat}
  {\frontmatter@RRAPformat{\produce@RRAP{*#1\href{mailto:#2}{#2}}}\frontmatter@RRAPformat}
  {}{}
}%
\makeatother
\nonstopmode
\begin{document}

\title[]{On the properties of Alfv\'enic switchbacks in the expanding solar wind: three-dimensional numerical simulations}

\author{Zade Johnston}%
 \email{johza721@student.otago.ac.nz}
 \affiliation{Physics Department, University of Otago, Dunedin 9010, New Zealand}
 \author{Jonathan Squire}
\affiliation{Physics Department, University of Otago, Dunedin 9010, New Zealand}
\author{Alfred Mallet}
\affiliation{Space Sciences Laboratory, University of California, Berkeley, CA 94720, USA}
\author{Romain Meyrand}
\affiliation{Physics Department, University of Otago, Dunedin 9010, New Zealand}

\date{\today}

\begin{abstract}
Switchbacks -- abrupt reversals of the magnetic field within the solar wind -- have been ubiquitously observed by Parker Solar Probe (PSP). Their origin, whether from processes near the solar surface or within the solar wind itself, remains under debate, and likely has key implications for solar wind heating and acceleration. Here, using three-dimensional expanding box simulations, we examine the properties of switchbacks arising from the evolution of outwards-propagating Alfv\'en waves in the expanding solar wind in detail. Our goal is to provide testable predictions that can be used to differentiate between properties arising from solar surface processes and those from the `\emph{in-situ}’ evolution of Alfv\'en waves in switchback observations by PSP. We show how the inclusion of the Parker spiral causes magnetic field deflections within switchbacks to become asymmetric, preferentially deflecting in the plane of the Parker spiral and rotating in one direction towards the radial component of the mean field. The direction of the peak of the magnetic field distribution is also shown to be different from the mean field direction due to its highly skewed nature. Compressible properties of switchbacks are also explored, with magnetic-field-strength and density fluctuations being either correlated or anticorrelated depending on the value of $\beta$, agreeing with predictions from theory. We also measure dropouts in magnetic-field strength and density spikes at the boundaries of these synthetic switchbacks, both of which have been observed by PSP. The agreement of these properties with observations provide further support for the Alfv\'en wave model of switchbacks.
\end{abstract}
\maketitle

\section{Introduction}

A striking observation by Parker Solar Probe \cite{Fox2016-lb} (PSP) during its passes of the Sun has been the presence of `switchbacks': abrupt reversals of the magnetic field within the solar wind \cite{Bale2019-sd, Kasper2019-tm, Dudok_de_Wit2020-pp, Horbury2020-bd, Mozer2020-qq, Laker2021-vo, Tenerani2021-fp}. Switchbacks exhibit primarily Alfv\'enic correlations between magnetic and velocity fluctuations with a nearly constant magnetic-field strength, implying (combined with electron strahl measurements \cite{Kasper2019-tm}) that they are local rotations of the magnetic-field vector. The mechanism that heats and drives the solar wind is still uncertain, with models of magnetically driven solar wind generally split into two categories: wave/turbulence driven (WTD) and reconnection/loop-opening (RLO) (see e.g., Ref.~\onlinecite{Cranmer2009-tr}). These models relate broadly to the heating of the solar wind by Alfv\'en waves and turbulence, or by energetic processes near the Sun such as magnetic reconnection. Given their significant energetic content, it is reasonable to hope that a better understanding of the origin of switchbacks may lead to further progress on these broader questions relating to the heating and launching of the solar wind itself.

Current theories of switchback formation fall into two classes, with a rough correspondence to these two solar-wind heating mechanisms. `\emph{Ex-situ}' mechanisms posit that impulsive events such as magnetic reconnection \cite{Drake2021-tb,Zank2020-vv,Schwadron2021-zt} and jets \cite{Sterling2020-oa} near the solar surface generate switchbacks. These mechanisms are mostly related to the RLO model of heating. In contrast, `\emph{in-situ}' mechanisms argue that switchbacks form due to processes within the solar wind itself, such as the development of non-linear Kelvin-Helmholtz instabilities \cite{Ruffolo2020-pi} across stream boundaries. These \emph{in-situ} mechanisms generally tie into the model of WTD heating.

Perhaps the simplest theory, which falls into the \emph{in-situ} class, is that switchbacks result from the evolution of Alfv\'en waves in the expanding solar wind. Alfv\'en waves are known to be common in the corona and solar wind \cite{Belcher1971-we,DePontieu2007-bn}, with their amplitude relative to the background magnetic field growing as they propagate outwards due to the expansion of the plasma \cite{Hollweg1972-fw,Heinemann1980-ji}. Ref.~\onlinecite{Squire2020-ji} used numerical simulations of expanding turbulence within the solar wind to show that switchbacks can form organically from an initial collection of outwards-propagating Alfv\'en waves. Switchbacks are also observed in the simulations of Ref.~\onlinecite{Shoda2021-ld}, where they analysed the evolution of outwards-propagating Alfv\'enic fluctuations within a magnetic flux tube stretching from the base of the solar corona out to 40 solar radii. Furthermore, Ref.~\onlinecite{Mallet2021-fh} (hereafter \mallet) examined the properties of non-linear Alfv\'en waves varying along one direction and gave a theoretical basis for how switchbacks arise from the constraints of constant magnetic-field strength and wave-amplitude growth in an expanding solar wind.

In this paper, we further examine the properties of switchbacks arising from the \emph{in-situ} evolution of Alfv\'en waves, and obtain a number of testable predictions that can be compared to switchback observations by PSP. We solve the locally isothermal MHD expanding box equations \cite{Grappin1993-jo,Grappin1996-bf,Dong2014-rz} approximating the evolution of a patch of solar wind outside the Alfv\'en point (where the Alfv\'en speed approximately equals the solar wind velocity) using high-resolution three-dimensional numerical simulations. We initialize these simulations with a collection of randomly phased, large-amplitude, outwards-propagating Alfv\'en waves with near constant magnetic-field strength (imagined to have propagated outwards from inside the Alfv\'en point), with this initial collection of waves displaying switchback-like features. The large amplitude of the fluctuations causes them to exhibit large magnetic field reversals, allowing switchbacks to evolve naturally. 

We investigate how the properties of these switchbacks depend on a range of parameters chosen to mimic those in the solar wind. A particular focus is the inclusion of the Parker spiral \cite{Parker1958-my}, i.e. a mean magnetic field with a non-radial component. We find that the asymmetry caused by this Parker spiral causes the magnetic field within switchbacks to be tangentially skewed and preferentially deflect towards the radial in the radial-tangential plane. This also causes the direction the peak of the magnetic field distribution to point in a different direction from the Parker spiral direction. The addition of the Parker spiral may also increase the fraction of switchbacks compared to a radial field, although this is dependent on the strength of turbulent effects. The $\beta$-dependent correlations between magnetic-field-strength and density fluctuations predicted by \mallet \ are also observed within these switchbacks. `Dropouts' in magnetic-field strength and density spikes at switchback boundaries with sharp changes in magnetic field and velocity components, shown to be a key property of switchbacks by Ref.~\onlinecite{Farrell2020-md}, are also observed in these simulations, suggesting many compressive properties of switchbacks can be understood from the \textit{in-situ} model Alfv\'enic model.

In our companion paper Ref.~\onlinecite{Squire_2022-yi} (hereafter \companion), we extend the theory of \mallet \ to the non-radial background field of the Parker spiral. We find that simple considerations arising from $\nabla\bolddot\B = 0$, the constancy of the magnetic-field strength, and the effects of expansion on wave amplitudes and wavenumbers allow one to infer a number of non-trivial effects that arise due to the Parker spiral. Combined with the results of \mallet, these results allow us to understand qualitatively most key results measured from the 3-D numerical simulations in this paper.

The results presented in this paper imply that the switchback properties that we measure arise naturally from the basic evolution of Alfv\'enic structures \emph{in-situ}; in our simulations, nothing is the product of solar-surface processes, since our initial conditions are simply a random collection of outwards-propagating waves. An example is the strong directional asymmetries of switchbacks within a Parker spiral, which we demonstrate straightforwardly can arise completely independently of any asymmetries in the source of Alfv\'en waves. These predictions can be tested against observations to help differentiate between the influence of \emph{in-situ} and \emph{ex-situ} processes on the properties of switchbacks within the solar wind.

An important complication of the Alfv\'enic \textit{in-situ} scenario, which unfortunately cannot be explored in detail in the standard expanding box model used here, is turbulence. As the plasma expands in our model, outwards-propagating waves reflect and generate inwards-propagating fluctuations, causing the development of reflection-driven turbulence \cite{Chandran2019-vr}. Outside the Alfv\'en point, the growth of the amplitude of fluctuations relative to the background magnetic field can stop if there is a strong enough turbulent cascade \cite{Chandran2009-vn}, meaning the growth of switchbacks may be stalled. The exact scalings for this turbulent behaviour remain highly uncertain, and such effects complicate predictions of the volume filling fraction and growth of switchbacks as a function of radius. However, we also argue based on previous works that conditions below the Alfv\'en point are extremely conducive for wave growth with or without turbulence, allowing for fluctuations to reach large amplitudes as they propagate outwards. 

\subsection{Outline}

In \cref{sec: theory}, we present the theory needed to understand how switchbacks evolve in the Alfv\'en wave model. We outline the expanding box model used in this paper (\cref{sub: ebm theory}), and give a summary of the results from our companion paper \companion \ (\cref{sub: aw sbs}), which will be compared to a number of diagnostics later in the work. The thorny issue of how turbulence may hinder the growth of switchbacks outside the Alfv\'en point is discussed in \cref{sub: turbulence theory}, although this remains uncertain.  This motivates brief consideration of switchbacks inside the Alfv\'en point in \cref{sub: inside RA}, allowing us to imagine large-amplitude outwards-propagating waves starting at the Alfv\'en point as our initial conditions. An overview of the simulations and numerical methods used in this paper is presented in \cref{sec:numerics}. Then, we investigate the properties of switchbacks generated by the evolution of Alfv\'en waves in \cref{sec: switchback properties}, which presents the evolution of the fraction of switchbacks and its dependence on turbulence within simulations (\cref{sub: sbfrac evo}), asymmetries arising due to the presence of a Parker spiral (\cref{sub: PS asymmetries}), and compressible properties of switchbacks (\cref{sub: compressible}). We conclude in \cref{sec: conclusion} with a summary of results and a discussion of how they relate to theory and observation.

\section{Theory}\label{sec: theory}

\subsection{Expanding plasma dynamics beyond the Alfv\'en point}\label{sub: ebm theory}

In this paper, we focus on the structures and properties of switchbacks arising from the evolution of large-amplitude Alfv\'en waves outside the Alfv\'en point, the heliocentric distance from the Sun $R_{\rm A}$, at which the Alfv\'en speed $v_{\rm A}$ approximately equals the solar-wind speed $U$. For $R\gg R_{\rm A}$, the solar wind has constant $U$, and its evolution can be described by the expanding box model (EBM) of Ref.~\onlinecite{Grappin1993-jo}. Here, the spherical expansion of a parcel of outwards-travelling plasma can be approximated, in the frame moving with the bulk solar-wind flow, by expansion perpendicular to the radial within a Cartesian frame. Aligning the $x$-axis with the outwards radial direction, the mass density $\rho$, flow velocity $\vel$, and magnetic field $\B$ evolve in this expanding frame as
\begin{subequations}
\label{eq:ebm_all}
\begin{align}
    \parfrac{\rho}{t}+\tildenabla\bolddot(\rho\vel)&=-2\frac{\dot{a}}{a}\rho\label{eq:ebm_density}, \\
    \parfrac{\vel}{t}+\vel\bolddot\tildenabla\vel &= -\frac{1}{\rho}\tildenabla\left(c^2_s\rho + \frac{\B^2}{8\pi}\right) + \frac{\B\bolddot\tildenabla\B}{4\pi\rho} - \frac{\dot{a}}{a}\bsf{T}\bolddot\vel\label{eq:ebm_momentum}, \\
    \parfrac{\B}{t}+\vel\bolddot\tildenabla\B &= \B\bolddot\tildenabla\vel - \B\tildenabla\bolddot\vel - \frac{\dot{a}}{a}\bsf{L}\bolddot\B\label{eq:ebm_induction}.
\end{align}
\end{subequations}
Here, $a(t)=1+\dot{a}t$ is the expansion parameter representing the growth of the perpendicular lengths of the frame, with $\dot{a}$ the constant expansion rate (due to constant $U$). The expansion parameter can be directly equated with $R(t)/R_0$, where the heliocentric distance $R(t) = R_0 + Ut$ for some initial $R_0$; this implies that $\dot{a}/a(t) = U/R(t)$. The gradient in the expanding frame is modified to $\tildenabla=(\partial_{x}, a^{-1}\partial_{y}, a^{-1}\partial_{z})$. The matrices $\bsf{T}=\textrm{diag}(0,1,1)$ and $\bsf{L}=\textrm{diag}(2,1,1)$ represent anisotropic `friction-like' terms due to conservation of angular momentum and magnetic flux during expansion.

Our simulations use a locally isothermal equation of state $P = c^2_s\rho$, with $P$ the thermal pressure and $c_s$ the speed of sound within the plasma. In this model, the temperature of the plasma is the same over entire domain at a given time. However, this temperature evolves with expansion as if the plasma was fully adiabatic, with the specific entropy $s = \ln (P/\rho^{5/3})$ being conserved. This, coupled with the locally isothermal equation of state, implies that $c_s\propto a^{-2/3}$, representing the cooling of the solar wind with expansion.

The EBM reproduces key scalings of quantities seen within the solar wind beyond the Alfv\'en point. For a given quantity $f$, we decompose it into its spatial mean (or background) part $\overline{f}$ and its fluctuating part $\delta f = f - \overline{f}$. Conservation of mass, angular momentum, and magnetic flux within the box force the background density, velocity, and magnetic field to scale as $\overline{\rho} \propto a^{-2}$, $\overline{u}_x \propto a^{0}$, $\overline{u}_{y,z} \propto a^{-1}$, $\overline{B}_x\propto a^{-2}$, and $\overline{B}_{y,z} \propto a^{-1}$. The Alfv\'en speed, $\va=\mB/\sqrt{4\pi\overline{\rho}}$ then scales as $v_{\rm A}=|\va| \propto a^{-1}$ for a radial background field.

These scalings also allow the Parker spiral to be captured within this model \footnote{The presence of strong azimuthal flows would invalidate our approximations by introducing rotational forces in the frame of the plasma. However, in the classic Weber-Davis model \cite{Weber1967-qs}, the rotation of the plasma scales as $1/R^2$, showing that such effects become negligible by large $R$ and should be unimportant to the overall dynamics outside $R_{\rm A}$.}. Following the scalings above, the Parker angle $\parkerangle$ (defined as the angle of the background field from the radial in the $xy$-plane) scales as
\begin{equation}
    \tan\parkerangle = \overline{B}_y/\overline{B}_x\propto a; \label{eq:parker_angle_evo}
\end{equation}
this shows that a background magnetic field with an initial non-zero non-radial component will rotate away from the radial as the box expands.

In contrast to the scalings of background quantities above, the normalized amplitude of Alfv\'enic fluctuations in the WKB regime with frequencies $\gg \dot{a}/a$ scale as
\begin{equation}
    \AdB \equiv |\delta\B|/|\mB| \propto a^{1/2}, \quad \Adu \equiv |\delta\vel|/\vamag \propto a^{1/2}\label{eq: amp db du}
\end{equation}
for a radial background field. For a Parker spiral with small initial angle $\Phi_0 \ll 1$, these approximately scale as $\AdB \propto a^{1/2}(1 + a^2\Phi^2_0)^{-1/2}$ (and similarly for $\Adu$). This implies that, once $|\parkerangle|\gtrsim45^\circ$, the normalized amplitude of Alfv\'enic fluctuations decreases with expansion instead of increasing (see \S III B of \companion \ for more details).

PSP measures quantities in the Radial-Tangent-Normal (RTN) coordinate system, where the radial (R) is the direction pointing from the Sun to PSP, the normal (N) is the direction of the component of the solar north direction perpendicular to R, and the tangential (T) is the direction orthogonal to R and N such that the coordinate system is right-handed. Throughout this paper, we identify the $x$-, $y$-, and $z$-axes with the R, T, and N directions respectively. This places the Parker spiral within the RT-plane, as observed by PSP.

\subsection{Formation of switchbacks from large-amplitude Alfv\'en waves}\label{sub: aw sbs}

In essence, the \emph{in-situ} Alfv\'enic theory of switchback formation is
based on the properties of large-amplitude Alfv\'en waves and how they grow in an expanding plasma. Here we summarize some conclusions from \mallet \ and our companion paper \companion, which is focused on the Parker spiral. We
will see  elements of each of these conclusions show up in the 3-D simulation analyses below.
A key idea is that
\begin{equation}
    P  = {\rm const.},\quad \rho= {\rm const.},\quad \B^2 = {\rm const.},\quad \delta \bm{u} = \pm\frac{\delta \bm{B}}{\sqrt{4\pi \rho}}\label{eq: nl aw solution}
\end{equation}
(where $P$ is the plasma's thermal pressure) is a non-linear solution to the compressible (non-expanding) MHD equations, which propagates along the mean field  $\mB$ at  the speed $v_{\rm A}$.
Our simulations below will be initialized with a random 3-D field that approaches \cref{eq: nl aw solution} (there is small residual $\B^2$ variation); such states are seen ubiquitously in the solar wind.
Given their propagation speed and other properties, these solutions
are the non-linear generalization of the linear MHD Alfv\'en wave, with the interesting
property that -- regardless of the perturbation amplitude $\delta \B$ -- they do not distort and form shocks (unlike, for example, sound waves or magnetosonic waves; Ref.~\onlinecite{Barnes1974}). 
However, the constraint $\B^2={\rm const.}$ is quite severe, since coupled with 
$\nabla\bolddot \B=0$, it leaves only one degree of freedom for the magnetic field. Our results 
are based on understanding how these coupled constraints ($\nabla\bolddot \B=0$ and $\B^2={\rm const.}$), as well as wave growth, lead to reversals in 
the field -- i.e., switchbacks -- for fields that vary only along one direction $\ph$. 
While the 1-D assumption is certainly not truly valid in any realistic plasma, we suggest -- supported by the results of our simulations below -- that most of the results apply more generally, with the  $\ph$ direction corresponding to the direction of fastest variation for some 3-D structure. Thus, for example, a 1-D field with $\ph$
 nearly perpendicular to $\mB$ relates to 3-D structures that are 
 extended in the $\mB$ direction compared to the perpendicular direction. 
 This rough correspondence is unsurprising: the importance of the $\ph$ direction in 1-D solutions arises because $\nabla\bolddot\delta \B=0$ implies $\ph\bolddot\delta \B=0$, 
 so the correspondence simply relies on the $\nabla\bolddot$ being 
 dominated by variation in $\delta \B$ along some particular direction. 
 An additional effect of importance is that expansion causes $\ph$ to rotate towards the radial direction, i.e., 
 structures to become more extended in the perpendicular direction.
 
 Some key ideas, which  each relate to observations or our simulations below, are as follows:\vspace{0.1cm}\\
(i) Switchbacks form preferentially for highly perpendicular structures, \emph{viz.}, those with $\ph\bolddot\mB/|\mB|\ll1$ (\mallet). This property is a simple consequence of the fact that a switchback requires $\delta \B$ to have a significant component $\delta B_\|$ in the direction of $\mB$, which is not possible if $\ph$ and $\mB$ are nearly aligned because $\ph\bolddot\delta \B=0$. This parallel component is approximately given by 
\begin{equation}
\delta B_\| / |\mB| \sim \textrm{min}\{\AdB^2, \AdB\sin\vartheta\}, \label{eq:singlewave_sb}
\end{equation}
where $\vartheta$ is the angle between $\ph$ and $\mB$. $\delta B_\| / |\mB|$ scales as $a^{1/2}$ for oblique wavevectors and as $a^{-1/2}$ for nearly parallel wavevectors, showing that switchback growth from 1-D waves decreases once enough expansion has occurred, even if $\AdB$ keeps growing.\vspace{0.1cm}\\ 
(ii) In an otherwise perfect Alfv\'enic solution (\cref{eq: nl aw solution}), expansion generates small $\B^2$ perturbations (\cref{eq: amp db du}). These 
 perturbations, which arise from the compressive flow needed to change the shape of $\delta\B$  as it grows in amplitude, are minimized for $\beta$ of order unity, depending on the obliquity of the wave (see \mallet \ fig. 1). As this occurs, the compressive polarization ratio $\xi \propto \delta (\B^2)/\delta(\rho)$ changes sign.\vspace{0.1cm}\\ 
(iii) Counter-intuitively, the Parker spiral can significantly enhance the formation of switchbacks due to expansion, for 1-D waves that are initially modestly oblique (as opposed to highly oblique), even though the normalized wave amplitude grows more slowly with a Parker spiral. This  occurs because the rotation of the mean-field can aid in making a wave more perpendicular (thus forming more switchbacks per \cref{eq:singlewave_sb})
 before the Parker spiral rotates past $\parkerangle\simeq 45^\circ$ and the normalized wave amplitude starts decreasing.\vspace{0.1cm}\\ 
(iv) In the presence of a Parker spiral mean field in the $xy$-plane, switchbacks should preferentially involve perturbations in $\delta B_y$ (tangential field deflections), rather than in $\delta B_z$ (normal field deflections). The reason  is simply that for a random collection of wavevectors $\ph$ that are preferentially radial (due to expansion), those with $\ph$ in the $\hat{\bm{z}}$ direction are on average more perpendicular to $\mB$ than when $\ph\sim \hat{\bm{y}}$ (then see point (i) above). Alfv\'enic field perturbations are largest in the $\ph\boldtimes \mB$ direction, thus suggesting $\delta B_y$ perturbations preferentially cause larger switchbacks.\vspace{0.1cm}\\
(v) Tangentially directed switchbacks with a Parker spiral are asymmetric, meaning they preferentially deflect the magnetic field towards the radial direction (specifically the $\overline{B}_x$ direction) to cause a switchback fluctuation. This occurs as a consequence of maintaining $\B^2={\rm const.}$ through a region where $B_y$ crosses through zero, which requires $\delta B_x {\rm sign}(\overline{B}_x)$ to increase rather than decrease through the field rotation that forms the switchback.\vspace{0.1cm}\\
(vi) Compared to a radial background field, switchbacks that form from 1-D waves in a Parker spiral are sharper and more intermittent. By this, we mean that they feature more sudden reversals in the field, but these reversals are spaced between longer quiet periods (even when starting from smooth initial conditions).\vspace{0.1cm}\\ 
(vii) As a direct consequence of points (v) and (vi), in a constant-$B$ field with a Parker spiral, the mode of the magnetic field direction (i.e., its most common direction) is significantly rotated away from the radial compared to its mean (i.e., the Parker spiral direction, which is the propagation direction of perturbations). In other words, in the presence of large fluctuations, a measurement of the Parker angle from the most common field direction will give an answer that is significantly larger than the true Parker angle.

 Out of these conclusions, we consider (i), (ii), (iv) and (v) to be the more important for turbulent, 3-D situations (see below). This is because these 
 conclusions relate primarily to the structure of $\delta \B$ fluctuations 
at a given time, coupled (except for case (i)) to the effects of expansion 
 changing $\AdB$ and biasing structures to be more extended in the perpendicular than the radial direction.
In contrast, conclusions (iii), (vi), and (vii) likely relate more specifically to the  way 
that single waves grow and how $\ph$ rotates compared to the mean field. These 
will be strongly modified by turbulence, which, as we now discuss, causes
both strong interactions between different wavevectors, and additional 
damping of $\delta \B$ perturbations.

\subsection{The influence of turbulence on switchbacks}\label{sub: turbulence theory}

Processes near the solar surface and corona generate mainly outwards-propagating $\bz^+$ fluctuations that travel with the solar wind. However, the solar wind is demonstrably turbulent, which requires non-linear interactions between $\bz^+$ and $\bz^-$ perturbations to develop. The question, then, is how the $\bz^-$ perturbations are generated within the solar wind. The Alfvén speed within the solar wind decreases with distance from the Sun, as the magnetic fields and density of the plasma decay to satisfy conservation of mass and magnetic flux. This speed gradient can be shown to act as a reflection term for $\bz^+$ fluctuations, generating $\bz^-$ perturbations and causing a turbulent cascade via the process of reflection-driven turbulence \cite{Velli1989-yp,Matthaeus1999-fb,Cranmer2005-qg, Verdini2010-yc,Chandran2019-vr}.

\subsubsection{Reflection-driven turbulence beyond the Alfv\'en point}

Reflection-driven turbulence can be captured in the EBM used in this paper, with \cref{eq:ebm_all} containing terms representing the reflection of waves. To show this, we decompose the velocity and magnetic field into their background and fluctuating parts, and assume a Sunward radial background field and no background flow; the small Parker spiral angles we consider in this paper should not significantly change the results of this discussion. We further assume no density fluctuations ($\delta\rho = 0$), and that $\rho$ and $\mB$ are spatially homogeneous solutions that satisfy \cref{eq:ebm_all}. The Elsasser variables are then defined as $\bz^\pm = \delta\vel \pm \delta\bm{b}$, where $\delta\bm{b} = \delta\B / \sqrt{4\pi\rho}$ is the magnetic field in velocity units.  Finally, we assume that the fluctuations are incompressible and perpendicular to the background field: $\tildenabla\bolddot\delta\vel = 0, \ \delta\vel\bolddot\mB = 0$, and $\delta\B\bolddot\mB = 0$. As $\mB$ is Sunward pointing, the $\bz^\pm$ variables represent outwards- and inwards-propagating fluctuations, respectively.

Using these assumptions, \cref{eq:ebm_all} can be written in a form that highlights this reflection of waves using the variables $\tilde{\bz}^\pm \equiv a^{1/2}\bz^\pm$:
\begin{align}
    \parfrac{\tilde{\bz}^+}{t} - \va\bolddot\tildenabla\tilde{\bz}^+ = - a^{-1/2}\tilde{\bz}^-\bolddot\tildenabla\tilde{\bz}^+ - a^{1/2}\rho^{-1}\tildenabla p_{\rm tot} - \frac{\dot{a}}{2a}\tilde{\bz}^-, \label{eq: z plus evo}\\
    \parfrac{\tilde{\bz}^-}{t} + \va\bolddot\tildenabla\tilde{\bz}^- = - a^{-1/2}\tilde{\bz}^+\bolddot\tildenabla\tilde{\bz}^- - a^{1/2}\rho^{-1}\tildenabla p_{\rm tot} - \frac{\dot{a}}{2a}\tilde{\bz}^+. \label{eq:z minus evo}
\end{align}
Here, the first term on the right-hand side represents the non-linear interactions between the perturbations that give rise to a turbulent cascade, and the pressure gradient term, where $p_{\rm tot} = c^2_s\rho + \B^2/8\pi$, enforces the incompressibility of the fluctuations ($\tildenabla\bolddot\bz^\pm = 0$). The final term represents the generation of perturbations via reflection due to expansion. For reference below, we define the rms amplitudes $z^\pm \sim (\overline{(\bz^\pm)^2})^{1/2}$ and $\tilde{z}^\pm = a^{1/2}z^\pm$.

Ref.~\onlinecite{Verdini2007-xe} and Ref.~\onlinecite{Chandran2009-vn} suggest a simple phenomenology for understanding the behaviour of the $\bz^+$ fluctuations due to turbulence, requiring two assumptions. First, outwards-propagating fluctuations are assumed to dominate and have large amplitudes compared to inwards-propagating fluctuations (i.e., $z^- \ll z^+$), allowing us to neglect the $\bz^-$ reflection term in \cref{eq: z plus evo}; and second, in \cref{eq:z minus evo} the driving due to reflection balances the non-linear damping of $\bz^-$ fluctuations \cite{Dmitruk2002-lp}. This leads to the scaling $\tilde{z}^- \sim (\dot{a}\lambda^+/2)a^{1/2}$, where $\lambda^+$ is a characteristic length scale of the $\bz^+\bolddot\tildenabla$ term that causes the turbulent damping of $\bz^-$ (Ref.~\onlinecite{Chandran2009-vn} assume that the length scales of non-linear interactions $\lambda^{\pm} \propto a$). One then inserts this into the $\bz^+$ equation to derive scalings for the evolution of $\tilde{z}^+$. Because $z^- \ll z^+$, we neglect the $\bz^-$ reflection term in \cref{eq: z plus evo} and obtain $\partial_t \tilde{z}^+ \sim -a^{-3/2}(\tilde{z}^-/\lambda^-)\tilde{z}^+$, where $\lambda^-$ is the characteristic length scale of the $\bz^-\bolddot\tildenabla$ term. Inserting the scalings for $\tilde{z}^-$ and assuming $\lambda^+ \sim \lambda^-$ gives
\begin{equation}
    \tilde{z}^+ \sim a^{-1/2}. \label{eq:z plus scaling}
\end{equation}
Writing \cref{eq:z plus scaling} in terms of the Elsasser variables, we obtain the scaling $z^+ / \vamag \sim \textrm{const.}$ for the normalized amplitudes of the fluctuations. This shows that strong non-linear interactions between $\bz^+$ and $\bz^-$ fluctuations can counter the effects of the WKB growth of normalized amplitudes (\cref{eq: amp db du}), stagnating the growth of $\bz^+$ fluctuations.

The strength of the turbulent cascade arising from the non-linear interactions between $\bz^+$ and $\bz^-$ fluctuations can be measured with the parameter
\begin{equation}
    \chi \equiv \frac{k_\perp z^+}{k_\| \vamag}, \label{eq:5chidef}
\end{equation}
which compares the strength of the non-linear interactions (proportional to $k_\perp z^+$) to linear effects (proportional to $k_\| \vamag$) for the $\bz^-$ fluctuations. The second assumption above, where the driving due to reflection balances the non-linear damping of $\bz^-$ fluctuations in \cref{eq:z minus evo}, requires $\chi \gtrsim 1$ so that non-linear effects dominate and the system can become strongly turbulent, damping fluctuations via an energy cascade. In the opposite regime, $\chi \ll 1$, the turbulence will instead be weak, which invalidates the argument used above.

\subsubsection{The effects of turbulent damping on switchback formation}

A rough estimation of $\chi$ can be obtained by using $z^+/\vamag \sim |\delta\B| / |\mB| = \AdB$, giving
\begin{equation}
    \chi \approx \AdB\frac{k_\perp}{k_\|}. \label{eq:chinought}
\end{equation}
As discussed above (point (i) in \cref{sub: aw sbs}), \mallet \ found that switchbacks form preferentially in oblique structures with $k_\perp \gtrsim k_\|$ ($\sin\vartheta \sim 1$ in \cref{eq:singlewave_sb}), which is a simple consequence of $\nabla\bolddot\delta\B = 0$. However, increasing $k_\perp / k_\|$ increases the value of $\chi$, via \cref{eq:chinought}. This then implies that the system will be strongly turbulent, and thus satisfy the scaling \cref{eq:z plus scaling}, with expansion-induced growth of $\AdB$ and $\Adu$ balanced by turbulent decay.

Ideally, one would like to study the growth of switchbacks in the solar wind starting from small-amplitude, nearly linear Alfv\'en waves, as thought to be released from the solar surface. To form switchbacks, such waves must:\vspace{0.1cm}\\
(i) Have their normalized amplitudes grow to reach $\AdB\sim 1$ and $\Adu\sim 1$;\vspace{0.1cm}\\
(ii) Be (at least modestly) oblique with respect to the background magnetic field, with $k_\perp \gtrsim k_\|$ (see \cref{eq:singlewave_sb});\vspace{0.1cm}\\
(iii) Start with $\chi \lesssim 1$, so that they can grow as in the WKB regime without significant energy decay due to a turbulent cascade.

However, the wave obliquity, $k_\perp / k_\|$, scales as $a^{-1}$ due to expansion, which causes $\chi$ to scale as $a^{-1/2}$. This implies that the three constraints on switchback formation above are incompatible: if we start with $\chi < 1$ and $z^+ / \vamag \ll 1$ at low altitudes, then $k_\perp \ll k_\|$ by the time $z^+ / \vamag \sim 1$. But, increasing the initial $k_\perp / k_\|$ to counter this effect means that $\chi$ will be initially $\gg 1$, which causes strong turbulent decay, no growth of $z^+ / \vamag$, and thus no switchback formation. This implies that, within the EBM, it is likely not possible to form switchbacks from initially low-amplitude waves (unless they are close to one-dimensional, as in \mallet \ and \companion).

At first sight, the above argument appears to invalidate the \emph{in-situ} formation of switchbacks from small-amplitude Alfv\'enic fluctuations propagating outwards from low altitudes. However, the EBM scalings, on which these arguments rely heavily, are valid only far beyond the Alfv\'en point in the constant-velocity expanding wind.  As we show below (\cref{sub: inside RA}), amplitude scalings in the sub-Alfv\'enic wind are, in contrast, highly conducive to the formation of switchbacks, even in the presence of turbulence. Finally, it is also worth noting that the exact scalings for this turbulent decay remain highly uncertain, with the predictions of the phenomenological model above decaying modestly faster than what is observed in simulations and within the solar wind \cite{Van_Ballegooijen2016-sp,Chandran2019-vr}. Further investigation of reflection-driven turbulence is needed. In addition, other physical effects may also be at play within the solar wind, such as the helicity barrier, which stops the turbulent cascade from reaching small scales \cite{Meyrand2021-ix, Squire2022-dm}, thus presumably halting the decay of $\bz^+$.

\subsection{Wave growth and scaling inside the Alfv\'en point}\label{sub: inside RA}

The estimates of the previous paragraph appear rather pessimistic for the \emph{in-situ} formation 
of switchbacks from small-amplitude fluctuations at the 
solar surface: in an expanding constant-$U$ wind, 
random 3-D waves will become turbulent if $\chi\gtrsim1$, 
in which case $z^+/\vamag$ may not grow at all as waves propagate outwards. Further,  it is not 
possible to reach $z^+/\vamag\sim 1$ with $k_\perp\gtrsim k_\|$ (as needed for switchbacks) while maintaining $\chi\lesssim1$, because $\chi\propto a^{-1/2}$ is a decreasing function of $a$ even for linear WKB waves. 
However, these scalings apply only to the super-Alfv\'enic wind where 
$U\sim {\rm const.} \gg \vamag$, for heliocentric radii $R\gtrsim R_{\rm A}$. In this section, we show
that for $R<R_{\rm A}$, amplitude scalings are, to the contrary, extremely conducive to the formation of switchbacks, either with or without turbulence. 
The arguments we make here are based on well-known and
understood scalings \cite{Hollweg1974-so,Heinemann1980-ji,Velli1993,Chandran2009-vn,vanBallegooijen2011}
that produce  reasonable  agreement with  observations \cite[e.g.,][]{Cranmer2012,Shoda2022}. 
Further, global flux-tube simulations, which correctly capture this physics, have already been shown to produce 
switchbacks starting from low-amplitude initial conditions \cite{Shoda2021-ld}.

As described in e.g., Refs.~\onlinecite{Chandran2009-vn,Chandran2019-vr}, the $\tilde{\bz}^\pm$ equations \eqref{eq: z plus evo}--\eqref{eq:z minus evo} in the EBM
are in fact the $U\gg \vamag$ limit of more general evolution equations that also apply for $R<R_{\rm A}$ (equations 2.19-2.20 of Ref.~\onlinecite{Chandran2019-vr}).
These assume the existence of a near-radial flux tube, with background radial field strength $B_0(R)$ and mass density $\rho(R)$, and use  wave-action conservation 
\cite{Heinemann1980-ji} to show that the `generalized wave-action variables,'
\begin{equation}
    \chg = \frac{1+\eta^{1/2}}{\eta^{1/4}} \bz^+,\quad \chf = \frac{1-\eta^{1/2}}{\eta^{1/4}} \bz^-,\label{eq: HO wave scalings}
\end{equation}
propagate unchanged in the absence of reflection and 
non-linear interactions. Here $\eta \equiv \rho/\rho_{\rm A}$, where $\rho_{\rm A}$ is the value of $\rho$ at $R=R_{\rm A}$, 
meaning $R\ll R_{\rm A}$ ($R\gg R_{\rm A}$) evolution is described 
by the limit $\eta\gg1$ ($\eta\ll1$). In addition, as shown by Ref.~\onlinecite{Hollweg1974-so,Barnes1974} the scaling \eqref{eq: HO wave scalings} applies to 1-D Alfv\'enic (constant-$B$) fluctuations of arbitrary amplitude, even once $\AdB\gtrsim1$. In the EBM limit (see below), $\chgf$ become the  $\tilde{\bz}^\pm$ defined in equations \eqref{eq: z plus evo} and \eqref{eq:z minus evo}, as expected. For comparison to 
these scalings, we define $a^2$ to be the cross-sectional 
area of a flux tube, meaning magnetic-flux and mass-flux conservation imply $B_0 \propto 1/a^2$ and $\rho U \propto 1/a^2$, respectively. This further 
implies that $\vamag(R)= \eta^{1/2}U $, $a\propto \eta^{-1/2}U^{-1/2}=\eta^{-1/4}\vamag^{-1/2}$, and $U\propto a^2 \vamag^2$, with the additional information that for reasonable solar-wind solutions outside ${\sim}2R_{\odot}$, $\vamag$ decreases monotonically with $R$, while $U$ and $a$ increase with $R$. In addition, we note that in the absence of turbulence, 
wavevectors scale with the Lagrangian frame as $k_\perp\propto a^{-1}$ and $k_x\propto (U+\vamag)^{-1}$ (where $k_x$ refers to the radial wavenumber). The scaling
for $k_x$ arises because the frequency of the wave remains constant as it propagates outwards\citep{Voelk1973a}, as for standard WKB theory\footnote{In the published version of this article in Physics of Plasmas, the scaling of $k_x$ is incorrectly stated as $k_x\propto U^{-1}$, leading also to incorrect statements regarding the scaling
of $\chi$ with $R$. An Erratum that corrects this error is now published \citep{Johnston2022-fu}, but we opted to update the 
text of this \texttt{arXiv} version directly.}.


Expanding the scaling \eqref{eq: HO wave scalings} in the limit $\eta\ll1$ and $U\sim {\rm const.}$ as applicable to $R\gg R_{\rm A}$, we obtain the expected EBM results from \cref{sub: ebm theory}: $a\sim \eta^{-1/4}$, $\rho\sim a^{-2}$, $B_0\sim a^{-2}$, $\vamag\sim a^{-1}$, giving  $\AdB\sim z^\pm/\vamag\sim \eta^{-1/4}\chgf\sim a^{1/2}\chgf$ (recall that $\chgf$ are constant in the absence of reflection and non-linearity).
The opposite limit, $\eta\gg1$ as applicable to $R\ll R_{\rm A}$ gives $z^\pm\sim \pm\eta^{-1/4}\chgf $, or 
\begin{equation}
    \frac{z^+}{\vamag}\sim \frac{a^{3/2}}{U^{1/4}}\chg \sim \frac{a}{\vamag^{1/2}} \chg.
\end{equation}
Because $U$ and $\va$ are, respectively, increasing and decreasing functions of $R$, this shows that $\AdB$ increases as $a^\alpha$ where $1<\alpha<3/2$. 
This is a much more rapid increase in amplitude than with constant expansion. 
Applying the same reflection-driven turbulence phenomenology discussed above (\cref{sub: turbulence theory}) but without the $\eta\ll1$ EBM assumption, one finds 
$\chg\propto \vamag^{1/2}$ (see equation (25) of Ref~\onlinecite{Chandran2009-vn}), implying $\AdB\propto a$, \emph{viz.,} the turbulent decay remains  too weak to counter the strong 
amplitude growth caused by expansion (indeed, strong wave growth is seen in 
detailed simulations of such turbulence \cite{vanBallegooijen2011,Perez2013-db}).
Thus, fluctuation amplitudes continue growing even in the presence of turbulence, unlike in the
EBM, thus potentially reaching $z^+/\vamag\gtrsim 1$ as needed for switchbacks, even if $z^+/\vamag\ll 1$ in the low corona.
However, it is worth noting that in sub-Alfv\'enic regions with $\vamag\gg U$, 
$k_x \sim \vamag^{-1}$ implies that $\chi \sim a^{-1/2}U^{1/4}\propto \vamag^{1/2}$, which usually decreases with $R$ above modest altitudes (for $R\gtrsim R_\odot$; see, e.g., Ref.~\onlinecite{Feldman1997}). This suggests that if perturbations start near the Sun in a weak, nearly linear regime ($\chi\ll1$) they are unlikely to become strongly 
turbulent via expansion-related processes (note that an earlier version of this 
article incorrectly concluded $\chi$ increased with radius\cite{Note2}). While this may 
have interesting consequences, further discussion is beyond the scope of this work.

Overall, we see that to study switchback formation from small amplitude waves requires a model that can capture large-amplitude fluctuation evolution 
for $R\lesssim R_{\rm A}$. The accelerating expanding box of Ref.~\onlinecite{Tenerani2017}
may be appropriate for this for future study, but is well beyond the scope 
of this work. Another option,  global flux-tube simulations, have already demonstrated 
that switchbacks can form under such conditions \cite{Shoda2021-ld}, but are computationally expensive, limiting the available resolution. 
Thus -- as discussed in the introduction and elsewhere -- we 
focus on the properties of large-amplitude Alfv\'enic switchbacks, as opposed to 
their growth and scaling with $a$.

\section{Numerical Methods and Simulations}\label{sec:numerics}

\begin{table}
	\centering
	\caption{Properties of simulations studied in this work. The parameters explored include the value of the plasma beta, $\beta_0$, and the approximate normalized amplitude of the Alfv\'enic fluctuations, $A_{\delta\B,0}$, at the start of the expanding phase. All simulations have $\dot{a}=0.5$ and are expanded up to $a=5$. A radial background magnetic field along the $x$-axis is imposed for all simulations except for \parkerhires, which has an initial non-radial $y$-component chosen such that the Parker angle $\parkerangle = -15^\circ$ at $a=5$.}
	\label{tab:simulation_table}
	\begin{tabular}{lccr} 
		\hline
		Name & Resolution & $\beta_0$ & $A_{\delta\B,0}$  \\
		\hline
		 \defaulthires & $1200^3$ & $0.35$ & 1 \\
         \parkerhires & $1200^3$ & $0.35$ & 1  \\
         \betalow & $800^3$ & 0.16 & 1   \\
         \betahigh & $800^3$ & 1.0 & 1   \\
         \chilow & $800^3$ & 0.3 & 0.2  \\
         \chihigh & $800^3$ & 0.5 & 1.4 \\
         \hline
		\hline
	\end{tabular}
\end{table}

\subsection{Numerical solution of the expanding box model}

To solve \cref{eq:ebm_all} of the EBM, we use the finite-volume astrophysical code \athenapp \cite{Stone2008-td,Stone2020-kv}. The HLLD Riemann solver of Ref.~\onlinecite{Mignone2007-lb}, modified to include the effects of expansion, is used as it is well suited to capture the sharp discontinuities commonly seen within switchbacks. A previous implementation of the EBM in \athenapp \ unfortunately led to small-scale numerical instabilities at large expansion factors \cite{Squire2020-ji}. In this paper, we use the variables \cite{Hellinger2005-ae,Bott2021-vp} $\rho' = \lambda\rho$, $\vel' = \BLambda^{-1}\bolddot\vel$, and $\B' = \lambda\BLambda^{-1}\bolddot\B$ in \athenapp, where $\BLambda=\text{diag}(1, a, a)$ and $\lambda \equiv \text{det }\BLambda = a^2$. Extensive numerical testing has shown the implementation of these variables to be more robust and stable, allowing for simulations with larger expansion factors. Further details of the modifications to the HLLD solver are given in the Appendix.

\subsection{Simulation parameters and initial conditions}

For this paper, we choose a small set of simulations (\Cref{tab:simulation_table}) to illustrate our key points. All simulations are initially in a cubic domain, with $L_x = L_\perp = 1$ (using $L_\perp$ to denote both $L_y$ and $L_z$). A background magnetic field $\mB$ is set in the $xy$-plane with $\overline{B}_x < 0$ and $|\mB|=1$ initially, as well as a uniform mass density $\rho=1$. All simulations have $\dot{a} = 0.5$ in order to be relevant to the outer scale of the solar wind turbulent cascade. The key parameters we vary across the simulations are the initial Parker spiral angle, the initial value of $\chi$, and the plasma $\beta$. We investigate how the evolution of switchbacks is affected by a non-radial background magnetic field in \parkerhires, where the initial value of $\overline{B}_y > 0$ is chosen (via \cref{eq:parker_angle_evo}) such that $\parkerangle = -15^\circ$ at $a=5$. This corresponds to the Parker angles observed by PSP, which typically sees $\parkerangle$ between $-10^\circ$ and $-20^\circ$. To showcase how the growth of fluctuations is affected by turbulent decay, we vary the initial value of $\chi \approx \AdB$ within the simulations. Here, we use $k_\perp / k_\| \approx L_x / L_\perp = 1$ to approximate the initial value of $\chi$ at the outer scales of the domain. All simulations have $\chi \approx 1$ initially, except for \chilow \ and \chihigh \ which have $\chi \approx 0.2$ and $1.4$. Finally, we investigate the dependence of compressible properties within switchbacks by changing the initial value of the plasma beta
\begin{equation}
    \beta \equiv 8\pi c^2_s \overline{\left(\rho/\B^2\right)} \label{eq:beta}
\end{equation}
in \betalow \ and \betahigh, which have $\beta = 0.16$ and $1$ initially. 

The box length $L_x$ corresponds to the physical length scale $2.4\times 10^6$ km, 
and an outwards-propagating Alfv\'en wave with parallel wavelength $L_x$ has frequency $8.5\times 10^{-5}$ Hz (equations 4-6 of Ref.~\onlinecite{Squire2020-ji} 
with $\Gamma_{\rm sim} = 0.5$). Assuming a constant radial wind speed of 350 km/s,
 a Parker angle of $\parkerangle = -15^\circ$ at $a=5$ corresponds to a heliocentric radial distance of $\approx 50 R_{\odot}$ (where $R_{\odot}$ is the radius of the Sun). Because $a$ scales linearly with distance from the Sun, the simulations start out approximately at the Alfv\'en point at $R_{\rm A} \approx 10 R_{\odot}$.

\subsubsection{Generating spherically polarized initial conditions}

\begin{figure*}
\begin{center}
\includegraphics[width=\textwidth]{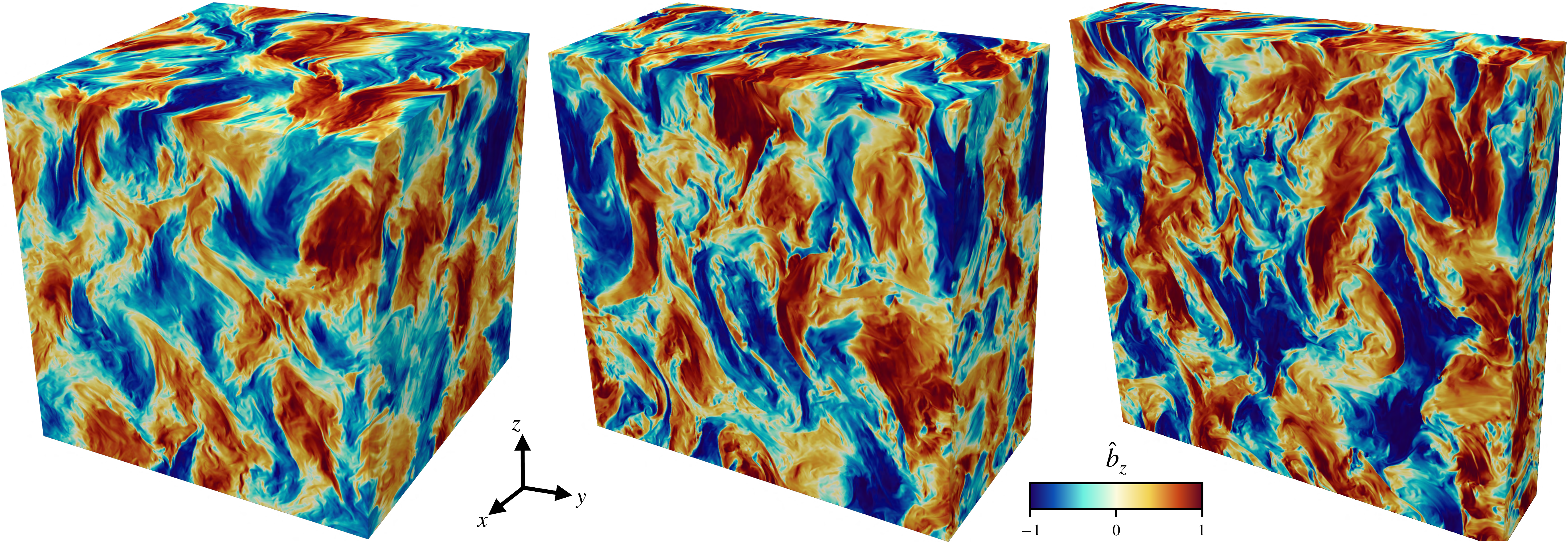}
\caption{Expansion of the \defaulthires \ simulation, showing the turbulent structure of the $z$-component of the magnetic field unit vector $\hat{\bm{b}} = \B / |\B|$. From left to right are snapshots at $a=1$, 2, and 5 (not to scale); structures appear to become sharper with expansion.}
\label{fig:turbulence_evo}
\end{center}
\end{figure*}

Waves with near constant magnetic-field strength $\B^2$ and Alfv\'enic correlations between the magnetic fields and velocity are commonly seen in the solar wind. These waves are a non-linear solution of the compressible MHD equations (\cref{eq: nl aw solution}), and are often called spherically polarized due to the constant-$\B^2$ constraint. To better mimic the conditions of solar wind, we wish to initialize the simulation in a spherically polarized state. A useful way to quantify this is the `magnetic compression' \cite{Squire2020-ji}:
\begin{equation}
    \CB \equiv \frac{(\B^2)_{\rm RMS}}{(\B_{\rm RMS})^2} = \sqrt{\overline{\left(\B^2 - \overline{\B^2}\right)^2}} \ \big/ \ \overline{\left|\B - \overline{\B}\right|^2}. \label{eq:CB2}
\end{equation}
$\CB$ is a measure of how the components of $\B$ are correlated to keep $\B^2$ constant. This is a non-linear effect that is relevant only for large amplitude waves. For example, when ${\AdB \ll 1}$, fluctuations $\delta\B$ perpendicular to the mean magnetic field $\mB$ with total magnetic-field strength $|\B|= |\mB + \delta\B|$ result in $|\B| / |\mB| = 1 + O(A^2_{\delta\B})$, which is constant to first order in $\AdB$; however, when ${\AdB \sim 1}$, perpendicular fluctuations alone will result in large fluctuations in $\B^2$. In a perfect spherically polarized Alfv\'en wave, the components of the magnetic field are correlated in such a way as to keep $\B^2$ precisely constant, causing $\CB=0$; this allows the magnetic compression to be used as a proxy for the degree of spherical polarization of the waves.

Near the Sun, the measured $\CB$ is small, with values of approximately $0.1 - 0.3$ seen in data from PSP (Chen, personal communication). This low magnetic compression is reflected in the near constant magnetic-field strength observed in switchbacks. This suggests that initializing simulations with a small $\CB$ would be preferable, to better mimic the conditions within the solar wind around the Alfv\'en point. 

However, it is extremely difficult to initialize a constant-$\B^2$ state across a 3-D simulation in general, as the magnetic field must also satisfy $\nabla\bolddot\B=0$ leaving one degree of freedom available to completely specify the magnetic field. Although methods to generate an initial constant-$\B^2$ state have been explored in other work \cite{Roberts2012-of,Valentini2019-oc, Squire2022-jt}, a simpler method is to let the system relax to a constant-$\B^2$ state by evolving in the non-expanding MHD regime.

The simulations looked at in this work are initialized with a collection of outwards-propagating, linear $\bz^+$ waves with random amplitudes and phases, to approximate an initially turbulent state. This causes $\CB \approx 1$ initially, due to there being no correlations between their components as there is no constraint keeping $\B^2$ constant. These waves can be thought of as a `superposition' of non-linear, spherically polarized Alfv\'en waves with constant-$\B^2$ and compressive fluctuations that cause $\B^2$ to change \cite{Barnes1974}. If this collection of waves is allowed to evolve without expansion, the compressible fluctuations rapidly dissipate by processes such as shocks, leaving behind a nearly constant-$\B^2$ Alfv\'enic state and reducing $\CB$ to values $\ll 1$. We note that this constant-$\B^2$ state inherits properties of initial collection of waves, such as its randomness; in other words, initializing the simulation with a different collection of waves gives rise to a different constant-$\B^2$ state, although its spectrum can of course change during the process. Ref.~\onlinecite{Squire2022-jt} present a different method for constructing large-amplitude 3-D constant-$\B^2$ states, which can give much smaller variation in $\B^2$ at the price of complexity, and may be of interest for initializing simulations in future work.

Based on this argument, we use the following method to generate near spherically polarized initial conditions in our simulations; namely, using the evolution of non-expanding MHD itself. The steps are as follows:\vspace{0.1cm}\\
(i) We initialize the simulations with a collection of outwards-propagating, linear $\bz^+$ waves using a sum of Fourier modes (i.e., waves with $\bz^- = 0$ or $\delta \vel = \delta \B / \sqrt{4\pi\rho}$), and are polarized like linear Alfv\'en waves in the $\bm{k}\boldtimes\mB$ direction. The waves are initialized with random amplitudes -- which follow a given energy spectrum $E(\bm{k})$ -- and random phases to approximate a turbulent initial condition.\vspace{0.1cm}\\
(ii) The collection of waves is then evolved without expansion for one Alfv\'en period. This causes the system to evolve towards a state with $\CB \approx 0.3$ as the waves decay and rearrange themselves.\vspace{0.1cm}\\
(iii) This low-$\CB$ state is then used as the initial condition for the expanding regime. Note that waves in this state already display some switchback-like features (as in \cref{fig:default_flyby}a below), as would be the case for fluctuations propagating from $R < R_{\rm A}$. 

The initial collection of Alfv\'en waves are given a Gaussian energy spectrum $E(k_x, k_\perp) \propto \exp\{-[(k_x - k_{x,0})^2+(k_\perp - k_{\perp,0})^2]/k^2_{w}\}$, where $k_\perp = (k^2_y + k^2_z)^{1/2}$. This was chosen as it gave higher switchback fractions than using other choices of initial energy spectra, a feature also seen in Ref.~\onlinecite{Squire2020-ji}, and likely related to a higher fraction of nearly perpendicular wavenumbers. The parameters $k_{x,0}=\kappa_\|(2\pi / L_x)$ and $k_{\perp,0}=\kappa_\perp(2\pi / L_\perp)$ set the centre of the Gaussian peak in $k$-space, and $k_w = 12 / L_\perp$ sets the width of the peak. To have initially large-scale fluctuations near the box scale, we set ${\kappa_\| = \kappa_\perp = 2}$. The decay of the waves during the non-expanding phase causes the fluctuation amplitude $A^2_{\delta\B}$ to also decay. Because of this, we initialize simulations with a larger normalized amplitude during the initial non-expanding phase, so that it decays to reach the values listed in \Cref{tab:simulation_table} at the beginning of the expansion.

\subsection{Evolution of global properties in simulations}\label{sub: global results}

\begin{figure}
\begin{center}
\includegraphics[width=\columnwidth]{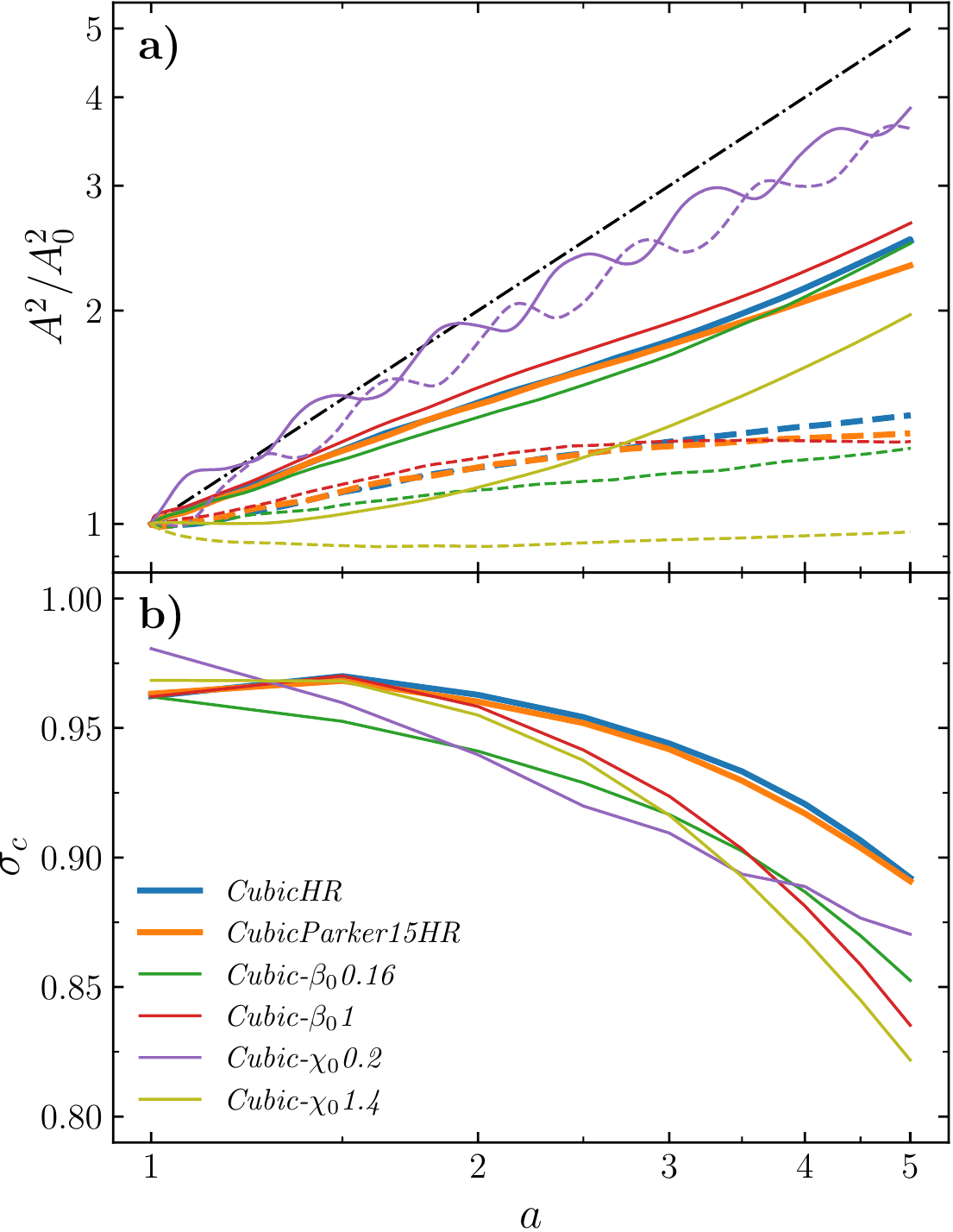}
\caption{Evolution of fluctuating quantities with expansion in the simulations listed in \Cref{tab:simulation_table}. Panel (a) shows the squared normalized fluctuation amplitudes $A^2_{\delta\B}$ (solid) and $A^2_{\delta\vel}$  (dashed), scaled by their initial value at $a=1$. In all simulations except for \chilow \ these normalized amplitudes grow slower than the expected WKB wave growth (dotted-dashed line) proportional to $a$, which is a consequence of the influence of turbulent decay on the growth of fluctuations. Panel (b) shows the decrease in normalized cross helicity $\sigma_c$ (\cref{eq:crosshelicity}) with expansion due to the growth of $\bz^-$ fluctuations from the reflection of $\bz^+$ fluctuations.}
\label{fig:amp evo}
\end{center}
\end{figure}

\begin{figure}
\begin{center}
\includegraphics[width=\columnwidth]{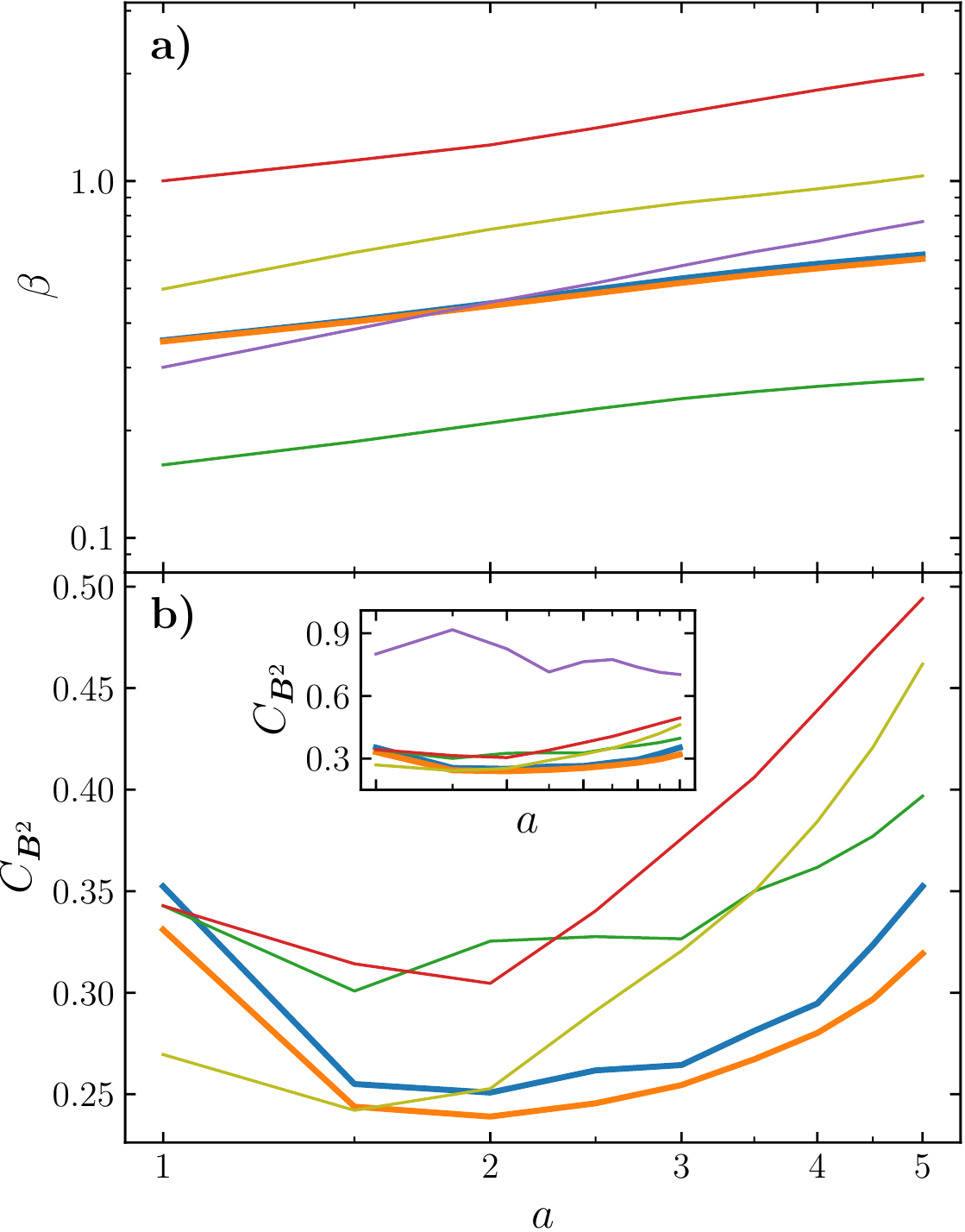}
\caption{Evolution of the plasma beta (a) and magnetic compression $\CB$ (b) with expansion, with line styles as in \cref{fig:amp evo}. The majority of simulations start with a small $\CB$, highlighting the near spherically polarized nature of fluctuations. The inset in panel (b) shows how the smaller initial amplitudes in \chilow \ gives rise to a larger $\CB$.}
\label{fig:beta evo}
\end{center}
\end{figure}

\Cref{fig:turbulence_evo} shows the emergence of turbulent structures with expansion in the \defaulthires \ simulation, showing the $z$-component of the magnetic-field unit vector $\hat{\bm{b}} = \B / |\B|$. Visually, these structures become sharper with expansion, a feature that is also seen in \cref{fig:default_flyby} below. \Cref{fig:amp evo} shows the evolution of fluctuating quantities with expansion within all simulations. The evolution of the normalized amplitudes $\AdB$ and $\Adu$ is shown in \cref{fig:amp evo}a, and is compared to the expected WKB growth of waves with expansion proportional to $a^{1/2}$. 
The normalized amplitudes of fluctuations within simulations with $\chi \gtrsim 1$ grow slower than the linear prediction, as is especially noticeable in \chihigh, which hardly grows at all. This is in reasonable agreement with the phenomenology in \cref{sub: turbulence theory}, where the balancing of the non-linear interactions and reflections of $\bz^\pm$ pertubations can cause amplitude growth to stagnate. In contrast, fluctuations in the \chilow \ simulation with their lower initial amplitudes are able to nearly follow the WKB prediction due to the reduced strength of non-linear interactions. 

We show the evolution of the normalized cross helicity
\begin{equation}
    \sigma_c = \frac{\overline{(\bz^+)^2} - \overline{(\bz^-)^2}}{\overline{(\bz^+)^2} + \overline{(\bz^-)^2}} \label{eq:crosshelicity}
\end{equation}
in \cref{fig:amp evo}b. This quantity is a key diagnostic of the properties of turbulence both within simulations and the solar wind, as there exist no non-linear interactions when $\sigma_c = \pm 1$. All simulations start out with $\sigma_c$ near 1, after decreasing slightly from $\sigma_c = 1$ during the non-expanding relaxation phase. The normalized cross helicity decreases further with expansion, due to the generation of $\bz^-$ fluctuations from the reflection of $\bz^+$ perturbations. Despite being in a weaker turbulent regime from the smaller wave amplitudes, the normalized cross helicity in \chilow \ also decreases with expansion; we suspect this is due to the evolution of non-WKB modes with $k_x=0$. 

In \cref{fig:beta evo}, we show the evolution of $\beta$ (\cref{eq:beta}) and the magnetic compression $\CB$ (\cref{eq:CB2}). Except for \chilow, all simulations start with $\CB \approx 0.3$ due to the initial non-expanding relaxation phase, showing that fluctuations approach a spherically polarized state with small fluctuations in $\B^2$. The relaxation of the system to this state is a non-linear effect, with simulations with greater initial amplitudes such as \chihigh \ reaching smaller values of $\CB$ compared to the low-amplitude \chilow \ (inset). This is because $\B^2$ is already nearly constant due to the small amplitudes of the fluctuations, and $\CB$ only measures the correlations between components. The magnetic compression in \defaulthires \ and \parkerhires \ is minimized compared to the \betalow \ and \betahigh \ simulations; this is discussed in more detail below in \cref{subsub: beta dependence}.

\section{Properties of Alfv\'enic switchbacks}\label{sec: switchback properties}

\begin{figure*}
\begin{center}
\includegraphics[width=\textwidth]{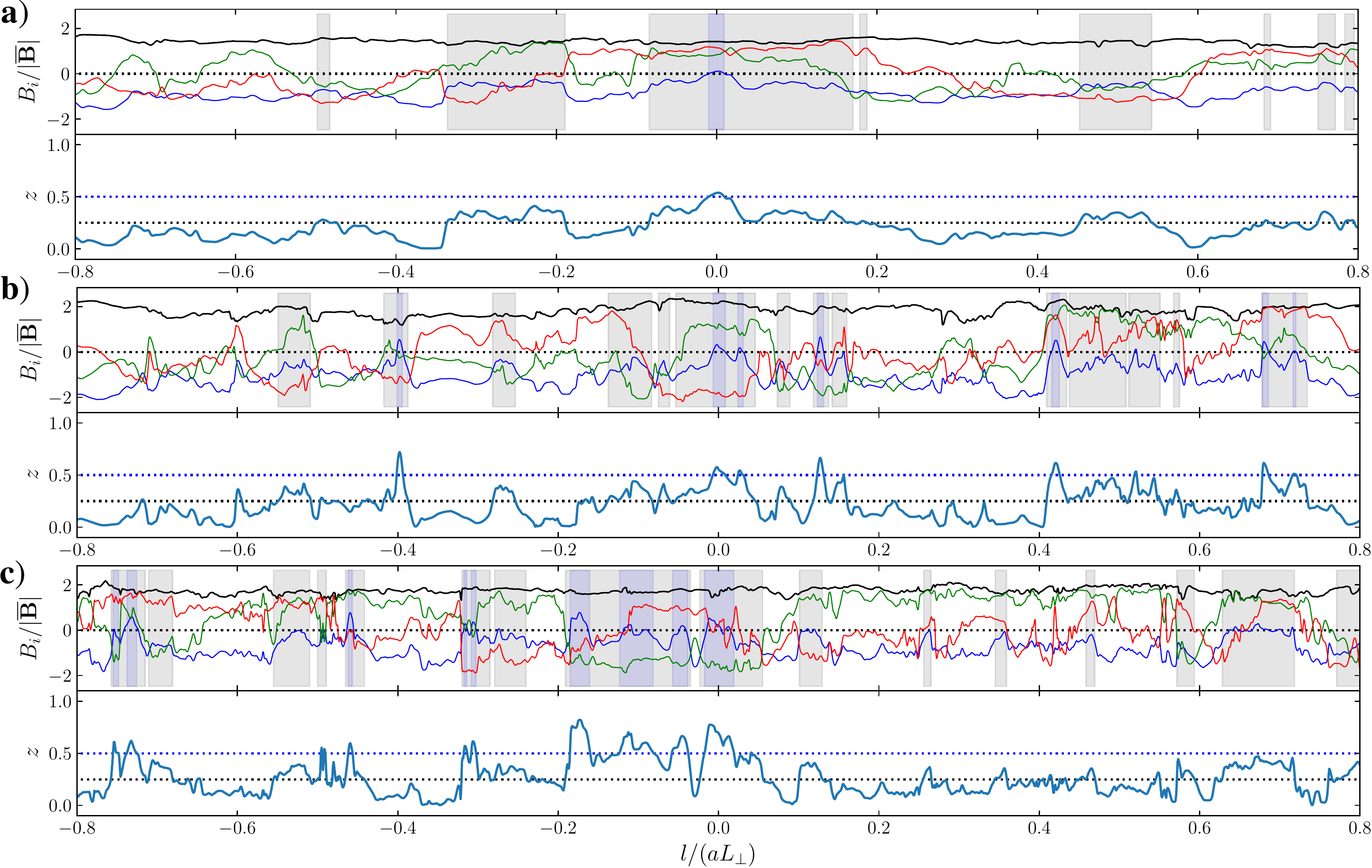}
\caption{Flybys through \defaulthires \ at $a=1$ (a) and $a=5$ (b), and \parkerhires \ at $a=5$ (c), showing the magnetic-field strength $|\B|$ in black and the components of $\B$ in blue ($B_x$), green ($B_y$), and red ($B_z$). All quantities are normalized to the background magnetic-field strength $|\mB|$. The bottom panels show the corresponding normalized deflection parameter $z$. Switchbacks with $z \geq 0.25$ are highlighted in grey, and those with $z \geq 0.5$ are highlighted in blue.}
\label{fig:default_flyby}
\end{center}
\end{figure*}

\begin{figure}
\begin{center}
\includegraphics[width=\columnwidth]{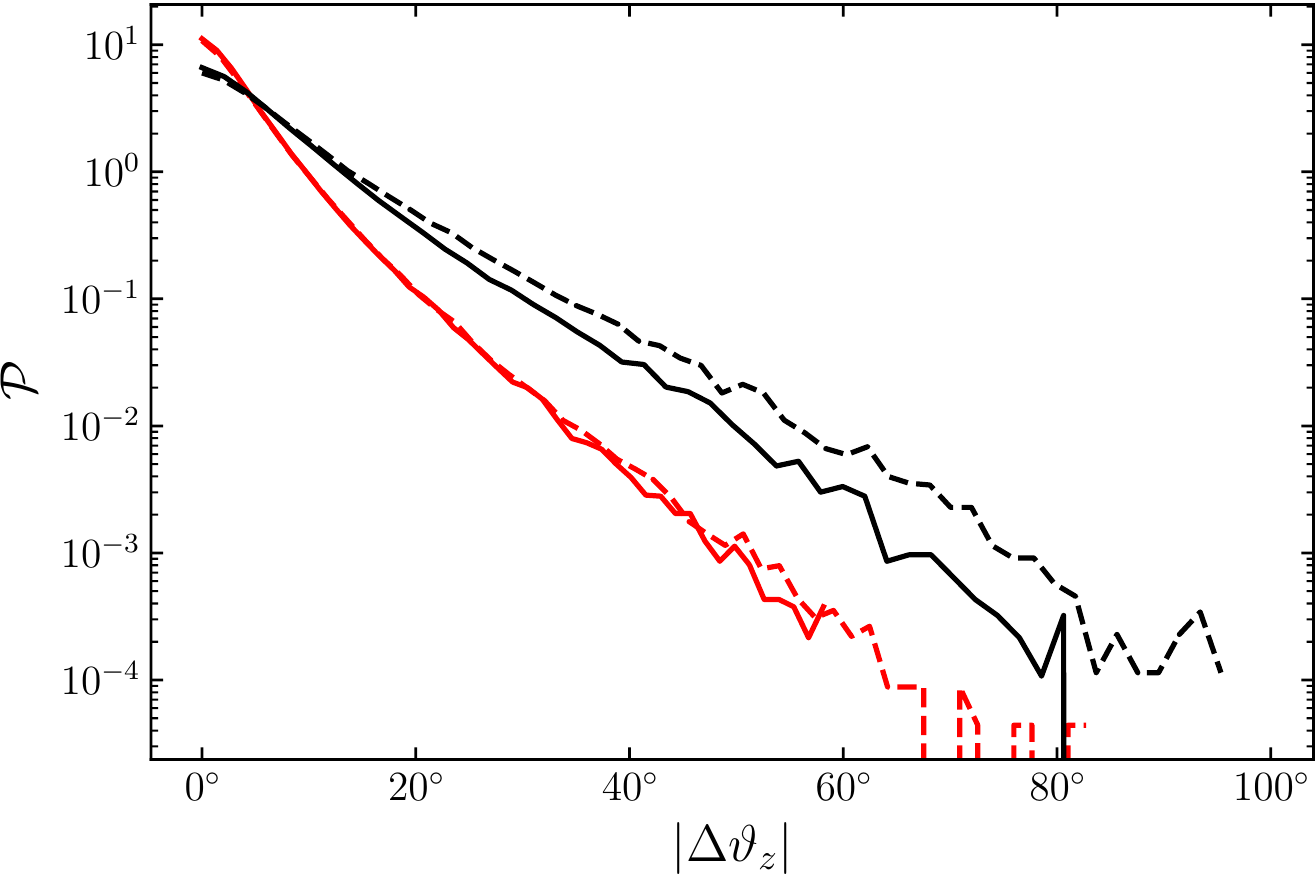}
\caption{PDFs of $|\Delta \deflectangle|$ (\cref{eq:dtheta_def}) from flybys in \defaulthires \ (solid) and \parkerhires \ (dashed), comparing the distributions at $a=1$ (red) and $a=5$ (black). Rotations become sharper with expansion in both simulations, with magnetic fields in a Parker spiral are exhibiting larger, sharper rotations than in the case of a radial background field. This feature is a prediction of the 1-D analysis of \companion.}
\label{fig:rotation_sharpness}
\end{center}
\end{figure}

In this section, we explore some more detailed properties of switchbacks within our simulation as they evolve with expansion. We study their magnetic field asymmetries and compressive properties, with the goal of comparing to basic theories of \mallet, \companion, and observations. We first define a switchback, a region where the magnetic field $\B$ has deflected from the background magnetic field $\mB$ by more than some threshold angle, via the normalized deflection parameter \cite{Dudok_de_Wit2020-pp}
\begin{equation}
    z \equiv \frac{1}{2}(1-\cos\deflectangle), \label{eq:z_deflect}
\end{equation}
with the deflection angle $\deflectangle$ given by
\begin{equation}
    \cos\deflectangle = \frac{\B\bolddot\mB}{|\B||\mB|}. \label{eq:costheta}
\end{equation}
Here, $z=0$ if the magnetic field and background magnetic field are parallel, and $z=1$ if they are antiparallel. We look at regions that satisfy $z\geq0.125$ to $z\geq0.75$ increasing in steps of $0.125$; $z=0.25, \ 0.5,$ and $0.75$ correspond to deflections from the background field of $60^\circ,\ 90^\circ,$ and $120^\circ$, respectively. These deflection angles were chosen to align with the observational definition of switchbacks, where they can be defined as deflections greater than $30^\circ - 45^\circ$ from the mean magnetic field \cite{Dudok_de_Wit2020-pp, Laker2021-vo, Laker2022-ij}.

\begin{figure*}
\begin{center}
\includegraphics[width=\textwidth]{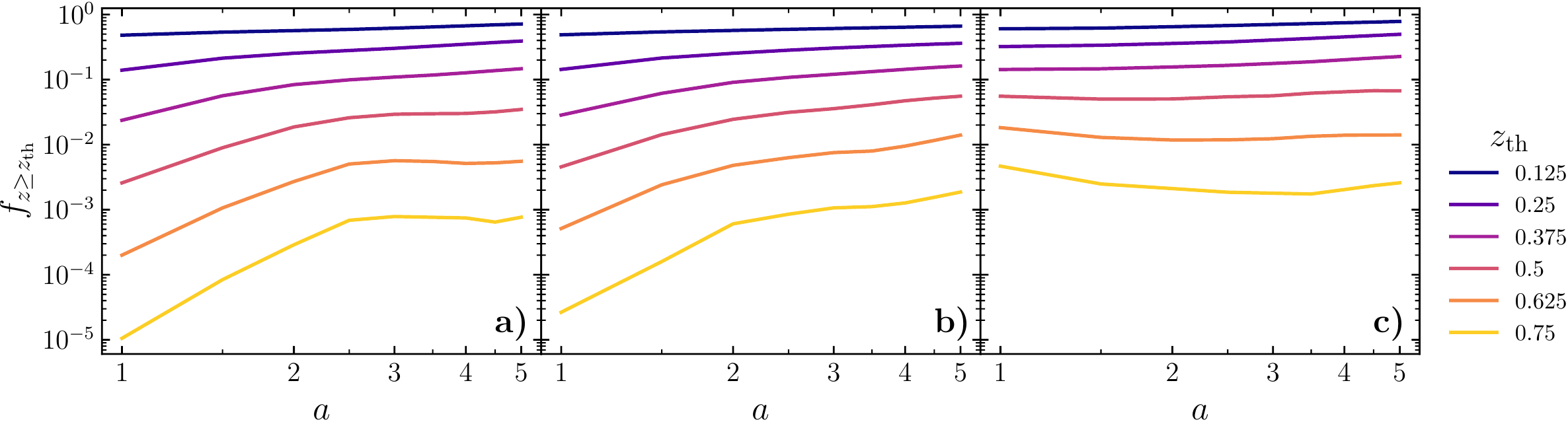}
\caption{Evolution of the switchback volume fraction $f_{z \geq z_{\rm th}}$ in \defaulthires \ (a), \parkerhires \ (b), and \chihigh \ (c), measuring the fraction of grid cells with $z$ greater than or equal to some threshold value $z_{\rm th}$. Switchbacks with larger deflections from the background field grow for longer with a Parker spiral than with a radial background field, due to the wavevectors of more waves staying oblique to the background field as they rotate with expansion. Panel (c) shows that waves with large amplitudes initially can give rise to a greater fraction of large magnetic field rotations, and that strong turbulent effects can stagnate the growth of these switchbacks.}
\label{fig:sbfrac_evo}
\end{center}
\end{figure*}

\Cref{fig:default_flyby} shows simulated flybys along the direction $(1, \sqrt{2}/2, \pi/8)$ through the \defaulthires \ simulation at $a=1$ (showing the large-amplitude initial conditions) and $a=5$ (\cref{fig:default_flyby}a and b), and the \parkerhires \ simulation at $a=5$ (\cref{fig:default_flyby}c), tracing the components of the magnetic field $\B$ and the magnetic-field strength $|\B|$. Due to Taylor's hypothesis of frozen-in flow, these flybys allow a crude representation of the structures PSP would see. Large rotations of the magnetic field are observed in both simulations, and switchbacks with $z \geq 0.25$ (grey) and 0.5 (blue) are common. The magnetic-field strength stays approximately constant throughout switchbacks, with only small fluctuations occurring, highlighting the near spherically polarized nature of the fluctuations. It is clear that the components of $\B$ are correlated to keep $\B^2$ constant, as expected from the small $\CB$ seen in \cref{fig:beta evo}b.

The rotations of the magnetic field appear to grow sharper with expansion, as shown by the steeper appearance of fluctuations in \cref{fig:default_flyby}b compared to in \cref{fig:default_flyby}a. We investigate whether this holds more quantitatively by calculating the change in $\deflectangle$ at each point $l$ along the flyby using
\begin{equation}
    \Delta\deflectangle(l) \equiv \langle \deflectangle \rangle_{\rm ahead} - \langle \deflectangle \rangle_{\rm behind}. \label{eq:dtheta_def}
\end{equation}
Here, $\langle \deflectangle \rangle_{\rm ahead}$ and $\langle \deflectangle \rangle_{\rm behind}$ denote an average of $\deflectangle$ over 5 grid points ahead and behind $l$. This averaging is done to ensure no grid-scale fluctuations are mistakenly identified as a rotation; we also found using larger averages over 10, 20, and 50 grid points made no significant difference to the results. \Cref{fig:rotation_sharpness} shows PDFs of $|\Delta\deflectangle|$ measured in flybys through \defaulthires \ and \parkerhires \ at both $a=1$ and $a=5$, with larger values of $|\Delta\deflectangle|$ corresponding to sharper rotations of the magnetic field. Magnetic fields tend to undergo sharper rotations with expansion, as seen in both \cref{fig:turbulence_evo} and \cref{fig:default_flyby}; this follows from the growth of fluctuation ampltiude with expansion (\cref{fig:amp evo}a). Furthermore, magnetic field rotations with a Parker spiral are more likely to be sharper compared to those with a radial background field. This property is expected based on the 1-D calculations of our companion paper \companion \ (point (v) in \cref{sub: aw sbs}), so is decent evidence of the applicability of 1-D arguments to complex 3-D fields. This is also an observationally testable prediction that could be studied in data.

\subsection{Switchback fraction evolution}\label{sub: sbfrac evo}

\Cref{fig:sbfrac_evo} shows the evolution of $f_{z\geq z_{\rm th}}$, the fraction of cells within the simulation containing switchbacks with $z$ greater than or equal to some threshold value $z_{\rm th}$. In the high resolution \defaulthires \ and \parkerhires \ simulations (\cref{fig:sbfrac_evo}a and b), we see the fraction of large deflections from the background field increase with the expansion of the box and the corresponding growth of the normalized amplitude of fluctuations. This increasing number of larger deflections from the background field with expansion seems to agree with observations \cite{Mozer2020-qq}. The addition of a Parker spiral causes the fraction of switchbacks to increase with expansion at large $a$, in contrast to the levelling-off of growth with a radial field. The evolution of three-dimensional switchbacks differs from the theory of 1-D Alfv\'enic solutions, which predicts that the switchback fraction decays once the wavevector reaches an angle $\lesssim 45^\circ$ from the mean field (\cref{eq:singlewave_sb}). For the switchback fraction to increase with expansion, there must be a mechanism that repopulates oblique modes that rotate towards the radial due to this expansion, as these modes  preferentially aid in the evolution of switchbacks (\mallet; point (i) of \cref{sub: aw sbs}). This repopulation is presumably provided by the evolution of the turbulence. As explained in \companion \ (also point (iii) of \cref{sub: aw sbs}), the rotation of the background field also aids in the evolution of the switchback fraction, as more wavevectors can remain oblique for longer as they rotate towards the radial.

However, this switchback growth does depend on the strength of turbulent effects, and so the detailed rate of switchback growth seen in \cref{fig:sbfrac_evo} remains only a qualified prediction of this model. In particular, both \defaulthires \ and \parkerhires \ start out with $\chi \approx 1$ which decreases with expansion, placing them on the boundary between the nearly linear WKB regime and strong turbulence, as seen in \cref{fig:amp evo}a. In contrast, the larger initial amplitude of waves in \chihigh \  causes amplitude growth to stagnate (see discussion in \cref{sub: turbulence theory}). \Cref{fig:sbfrac_evo}c shows that in this case, the switchback fractions are relatively constant with expansion. Interestingly, this means that as well as hindering the growth of switchbacks by stopping the growth of $\AdB$, turbulence cannot also destroy them even though the expansion naturally drives eddies to become more parallel. 

The dependence on the evolution of switchbacks on turbulence demonstrates two key points. First, in order to make detailed observational comparisons to switchback fraction evolution, it is crucial to understand the evolution of $\chi$ in the solar wind (as it could vary between streams). Second, it supports the idea of the growth of the normalized amplitude of perturbations is the key factor in the evolution of switchbacks: in this model, if the amplitude does not grow, the switchback fractions remain constant. It is also worth noting that the fraction of switchbacks with $z\geq 0.5$ in \chihigh \ is very high, $f_{z\geq0.5}\approx 5-7\%$, which is similar to that observed by PSP \cite{Bale2019-sd}. This demonstrates that 3-D Alfv\'enic states can exhibit very high switchback fractions, so long as the fluctuation amplitude is sufficiently high.

Similar trends are seen in lower resolution simulations, including a number of other tests not presented here, although in general $f_{z\geq z_{\rm th}}$ is a decreasing function of resolution (as also noted by Ref.~\onlinecite{Squire2020-ji} and Ref.~\onlinecite{Shoda2021-ld}). This is because the higher resolution reduces the effect of numerical dissipation in dampening out the quick and sharp changes typical of switchbacks, allowing for greater switchback fractions.

\subsection{Asymmetry of switchback magnetic-field deflections}\label{sub: PS asymmetries}

With a radial background field in the EBM, by symmetry, there should be no preferred direction for the magnetic field to deflect in. Including a non-radial component to the magnetic field (i.e., the Parker spiral) will break this symmetry, introducing a bias to these deflections. Here, we investigate asymmetries within switchbacks caused by including the evolution of a Parker spiral. 

\begin{figure}
\begin{center}
\includegraphics[width=\columnwidth]{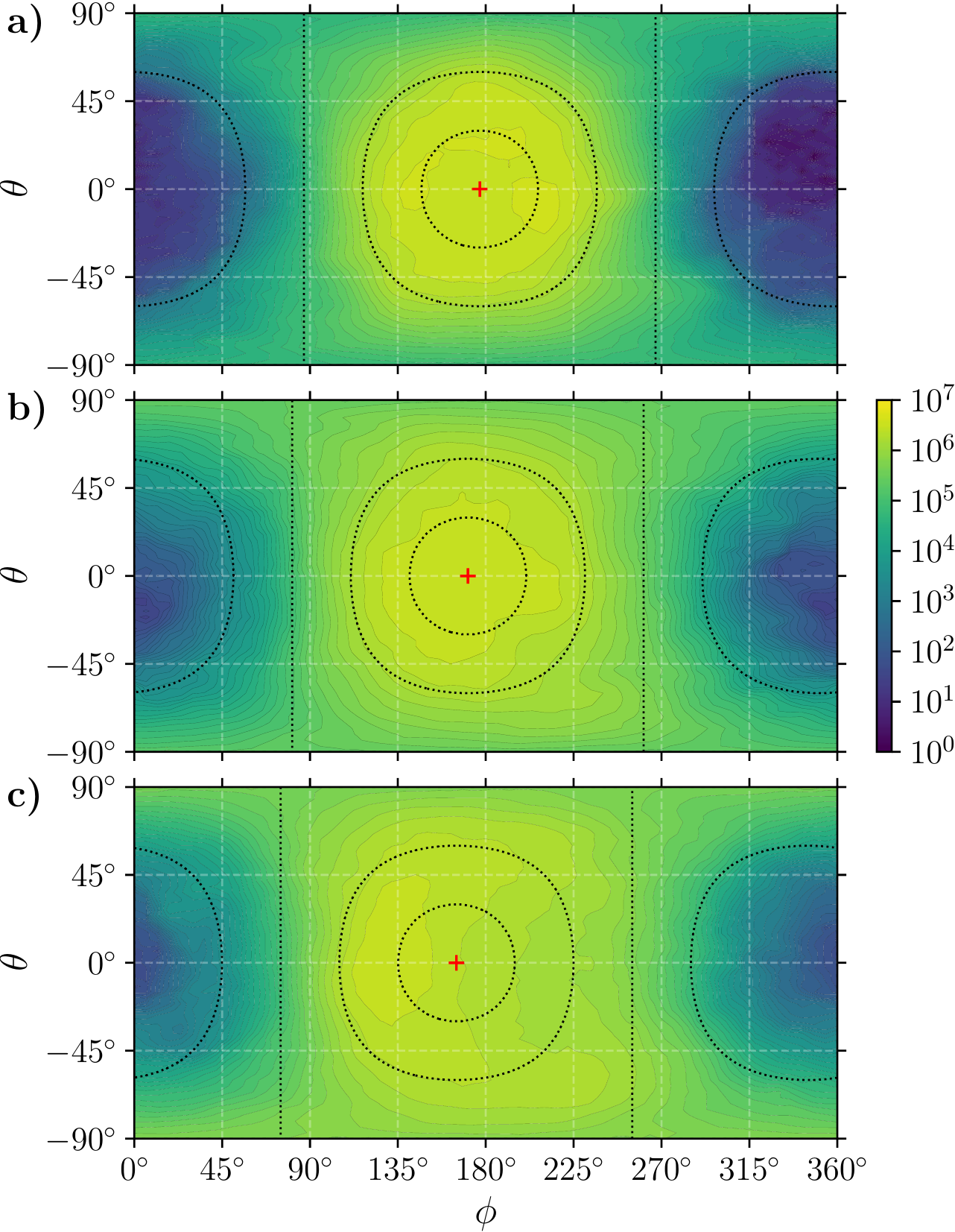}
\caption{2-D histograms of the angular deflections $\phi$ and $\theta$ (\cref{eq:deflection_angles}) of the magnetic field measured over the entire box in \parkerhires \ at $a=1$ (a), $a=2$ (b), and $a=5$ (c). The position of the mean magnetic field is shown by the red cross, with an angle of $\parkerangle = -6^\circ$ and $-15^\circ$ in (b) and (c); an angle of $\phi=180^\circ$ represents a Sunwards-pointing magnetic field. Dotted lines represent contours of $z = 0.07$, 0.25, 0.5, and 0.75, corresponding to deflections of $30^\circ, \ 60^\circ, \ 90^\circ$, and $120^\circ$ from the background field. The distribution shows a slight tangential bias for small deflections from the Parker spiral in b); however, the peak of the distribution shifts further towards the $+$T direction ($\phi=90^\circ$) compared to the background field at later times (c). In addition to this, a preference for deflections with $z \geq 0.25$ to point towards the $-$T direction ($\phi=270^\circ$) is also seen in (b) and (c), as shown by the slower drop-off in contours compared to the $+$T direction.}
\label{fig:B deflections}
\end{center}
\end{figure}

\begin{figure*}
\begin{center}
\includegraphics[width=\textwidth]{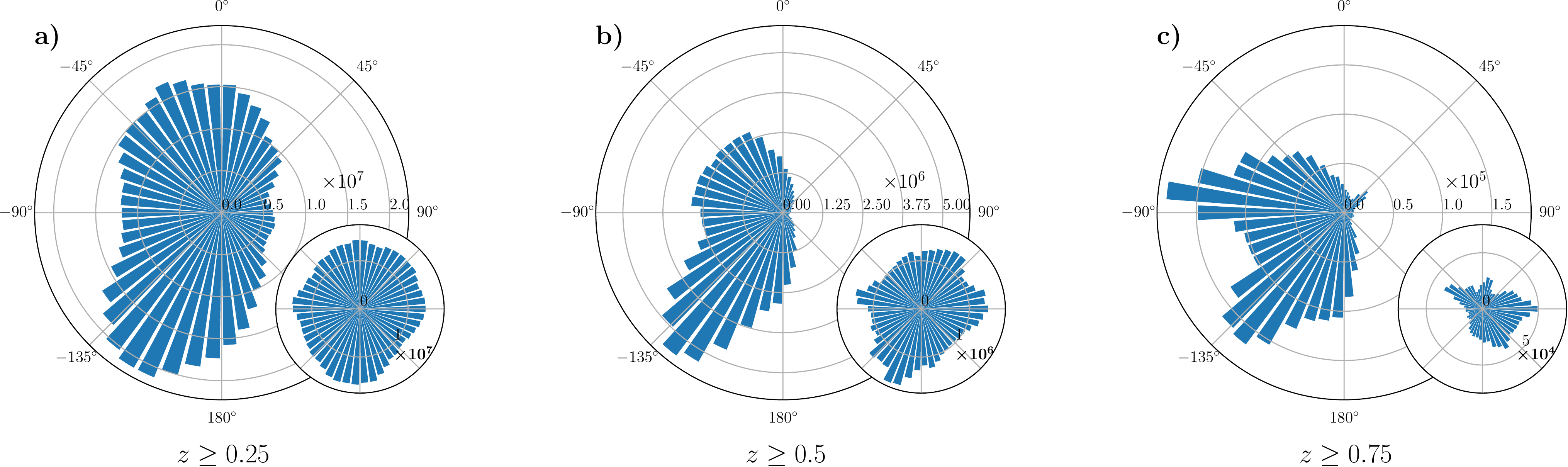}
\caption{Histogram of clock angles of all magnetic field vectors within regions of magnetic-field deflections satisfying $z\geq0.25$, $0.5$, and $0.75$ (a-c respectively) in \parkerhires \ at $a=5$. A clear deflection is shown towards the $-\rm T'$ direction ($-90^\circ$), with larger deflections showing a smaller spread in direction. Insets show corresponding distributions at $a=5$ in \defaulthires \ with a radial background field, which are nearly isotropic.}
\label{fig:clockangle_a5}
\end{center}
\end{figure*}

\begin{figure}
\begin{center}
\includegraphics[width=\columnwidth, height=8cm, keepaspectratio]{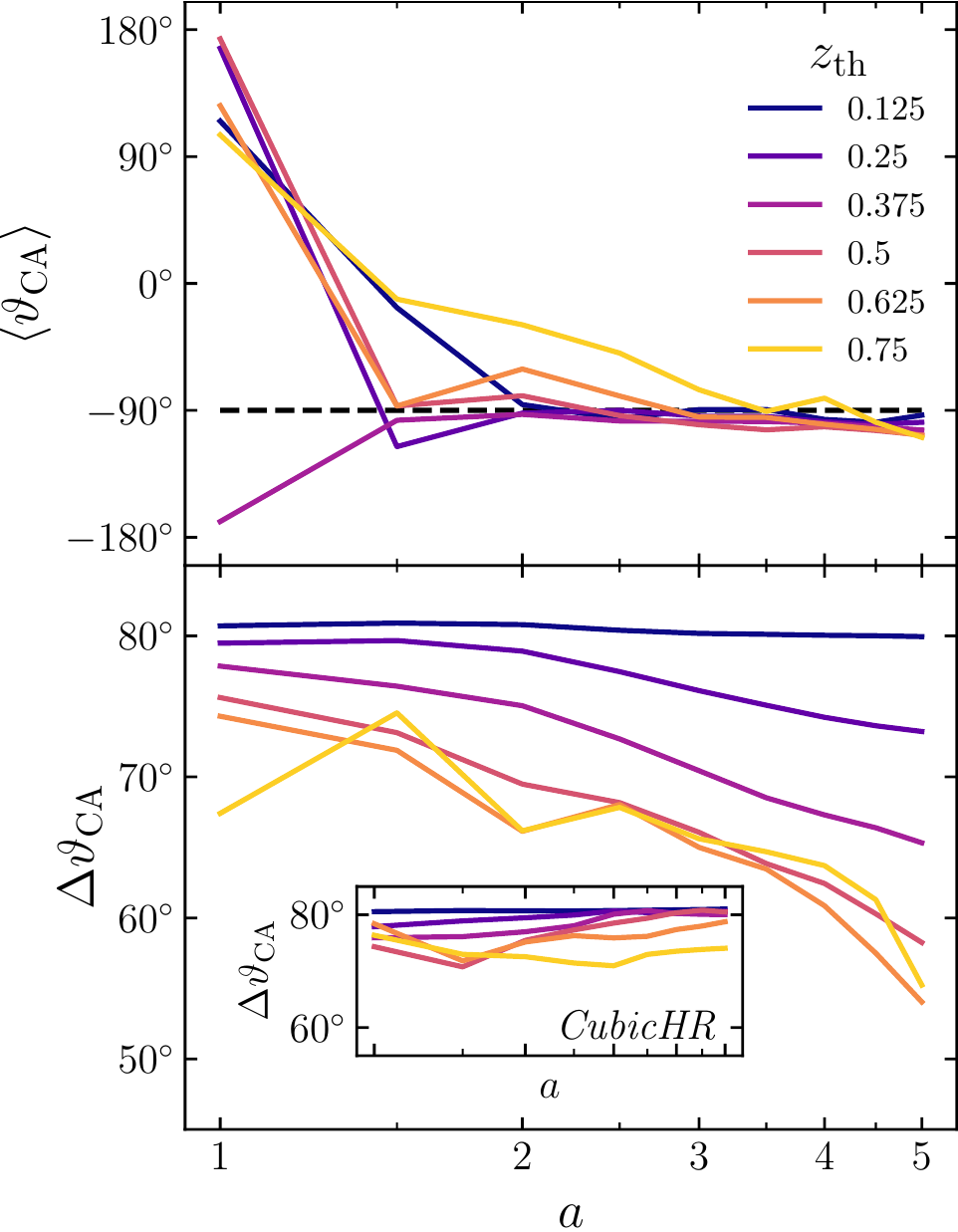}
\caption{Evolution of the mean $\langle \clockangle \rangle$ and angular spread $\Delta \clockangle$ of the clock angle distributions measured in \parkerhires. The distributions clearly centre around the $-\rm T'$ direction ($-90^\circ$, dashed black line) with expansion, with larger deflections more likely to point along $-\rm T'$ as shown by the decreasing angular spread. The inset shows the evolution of the angular spread of clock angle distributions measured in \defaulthires, showing that they are not as focused in a given direction compared to those with a non-radial background field.}
\label{fig:clockangle_evo}
\end{center}
\end{figure}

\subsubsection{Angular deflection distributions}\label{subsub: B deflections}

In their analysis of switchback observations by PSP, Ref.~\onlinecite{Dudok_de_Wit2020-pp} found that angular deflections of the magnetic field were nearly isotropic with respect to the Parker spiral. To see whether the magnetic fields in our simulations share this property, we plot the evolution of 2-D histograms of the angles
\begin{equation}
    \phi = \arctan\left(\frac{B_y}{B_x}\right), \quad \theta = \arcsin\left(\frac{B_z}{B}\right)\label{eq:deflection_angles}
\end{equation}
for every magnetic field vector within the \parkerhires \ simulation in \cref{fig:B deflections}. These are the angle from the radial in the RT-plane and the elevation out of the RT-plane towards the normal, respectively. We bin the angles of every magnetic field vector in a uniform grid in $\phi$ and $\theta$. To compensate for the shrinking of areas near the poles when these uniform grids are plotted on a sphere, we use a weighting factor of $1/\cos\theta$.

\Cref{fig:B deflections}a and b show the distributions at $a=1$ and $2$. Here, they are roughly centred on the Parker spiral, which itself is nearly radial (with $\parkerangle \approx -3^\circ$ and $-6^\circ$, respectively); these distributions are similar to angular distributions taken from \defaulthires, with its purely radial background field. Although nearly isotropically distributed around the Parker spiral initially at $a=1$, small deflections from the Parker spiral initially show a slight tangential bias, as shown by the elongated distribution for deflections with $z \leq 0.25$ in \cref{fig:B deflections}b; however, larger deflections become near-isotropically distributed. The distribution changes as the background field rotates further from the radial. An interesting feature at $a=5$ in \parkerhires \ (\cref{fig:B deflections}c) is the shift in the peak of the distribution towards $\phi = 90^\circ $ (the $+$T-direction) compared to the Parker spiral direction for deflections with $z \leq 0.25$. This shows that the direction of the mean field does not align with the most probable direction of the magnetic field. However, the larger counts of deflections with $z \geq 0.25$ towards $\phi = 270^\circ$ than $90^\circ$ suggests that the magnetic field within these switchbacks preferentially point in the opposite direction (towards the $-$T direction). More generally, the distributions in \cref{fig:B deflections} show that the Parker spiral causes strong asymmetry in switchback deflections purely as a result of expansion and field rotation, without requiring any asymmetry of the source (in our case the initial conditions of the simulations).

\subsubsection{Switchback clock angle}\label{subsub: clock angle}

An alternative measure of the direction of deflection of magnetic-field vectors within a switchback is the `clock angle' \cite{Horbury2020-bd}. For each magnetic field vector $\B$ within a switchback region, we project it onto the plane containing the N direction perpendicular to the background field $\mB$. This plane can be thought of as the TN-plane (or equivalently the $yz$-plane) rotated such that it is perpendicular to $\mB$; we denote the rotated tangential direction as $\rm T'$. The clock angle $\clockangle$ of the projected vector $\B_\text{proj}$ is then its angle measured clockwise from the +N-axis, given by
\begin{equation}
    \clockangle = \arctan\left(\frac{B_{\text{proj}, \rm T'}}{B_{\text{proj}, \rm N}}\right), \label{eq:clockangle}
\end{equation}
where $B_{\text{proj}, \rm T'}$ and $B_{\text{proj}, \rm N}$ are the $\rm T'$ and N components of $\B_\text{proj}$. A clock angle of $0^\circ, \ 90^\circ, \ 180^\circ,$ and $-90^\circ$ corresponds to the $+$N, $+\rm T'$, $-$N, and $-\rm T'$ directions, respectively.

\Cref{fig:clockangle_a5} shows polar histograms of the clock angle of magnetic field vectors inside switchbacks satisfying $z \geq z_{\text{th}}$ for $z_{\text{th}} = 0.25, \ 0.5$, and $0.75$ in the \parkerhires \ simulations at $a=5$, with similar histograms from \defaulthires \ at $a=5$ shown as insets. These histograms are related to the deflections in \cref{fig:B deflections}, and can be computed from the sum of all vectors lying outside the corresponding contour of $z$ (dotted lines in \cref{fig:B deflections}; in this way the information in \cref{fig:clockangle_a5} is a subset of that in \cref{fig:B deflections}). Magnetic field vectors with deflections along or near $\theta = 0^\circ$ in \cref{fig:B deflections} will have $\clockangle \approx \pm 90^\circ$ (with the sign depending on the direction of deflection from the Parker spiral), while those with $\theta \approx 90^\circ \ (-90^\circ)$ will have $\clockangle \approx 0^\circ \ (180^\circ)$.

The addition of a non-radial background magnetic field in \parkerhires \ causes magnetic-field vectors within switchbacks to evolve with a preferential deflection along or near the $- \rm T'$ direction, with this deflection along $- \rm T'$ becoming more pronounced for switchbacks with larger deflections from the background field. This is in contrast to switchbacks with a purely radial field in \defaulthires, which show no preferred deflection direction, as must be the case by symmetry. 

\Cref{fig:clockangle_evo} shows the evolution of the average clock angle $\langle \clockangle \rangle$ and angular spread $\Delta \clockangle$ of the clock angle distributions for the \parkerhires \ simulation. These quantify the average directional asymmetry of the distribution, as well as a measure of how focused the distribution is around this average direction. The clock angle distribution quickly becomes centred around the $-\rm T'$ direction ($\clockangle = -90^\circ$), with the angular spread of the distributions decreasing with increasing $z$. This shows that the clock angle of switchbacks with larger deflections from the background field are focused around the $-\rm T'$ direction (as seen in \cref{fig:clockangle_a5}). This contrasts with the \defaulthires \ simulation with just a radial mean magnetic field, with the inset in \cref{fig:clockangle_evo} showing a consistently higher angular spread of the distributions compared to those with a Parker spiral, as must be the case since the distributions inset in \cref{fig:clockangle_a5} are quasi-isotropic.

These deflections are in or near the RT-plane containing the Parker spiral, and points in the direction towards the radial component of the background field $\mB$. To confirm this result, we ran a lower resolution simulation with the Parker spiral reaching $\parkerangle = 15^\circ$ at $a=5$ with $\overline{B}_y < 0$, so that the background magnetic field rotates in the opposite direction as the box expands. Here the deflections were centred around the $+\rm T'$ direction, again pointing in the direction towards the radial component of $\mB$, showing that this is a robust effect of introducing a non-radial background magnetic field.

\subsubsection{Parker spiral: Discussion}

The results above show that the Parker spiral introduces a clear asymmetry into deflections of the magnetic field, both globally and inside switchbacks. The presence of a background field with even a small non-radial component, as in \cref{fig:B deflections}a and b, causes the distribution of small deflections from the mean to elongate more along the tangential direction compared to when a radial field is used. These results seem to roughly correspond with observations \cite{Dudok_de_Wit2020-pp,Horbury2020-bd,Laker2022-ij}, where preferential tangential deflections of the magnetic field are seen. 
As the mean field rotates further from the radial with expansion, the most probable direction that magnetic fields point in shifts even further towards the tangential (\cref{fig:B deflections}c). This shows that if this strong deflection is also true within data from PSP, care may need to be taken for the method of averaging the magnetic field in finding the direction of the Parker spiral.

\Cref{fig:B deflections} also shows the angular distribution of magnetic fields within switchbacks with large $\deflectangle$. In \cref{fig:B deflections}c, larger counts of magnetic field deflections with $z\geq0.25$ are seen towards $\phi = 270^\circ$ over $\phi = 90^\circ$, corresponding to a preferred deflection in the $-$T over the $+$T direction. These deflections correspond to the asymmetric distribution of deflections seen in the clock angle distributions in \cref{fig:clockangle_a5} and \cref{fig:clockangle_evo}.

This asymmetrical nature of tangential magnetic-field deflections within a switchback is a robust property of the Alfv\'en wave model of switchbacks, with the simulation results shown here matching with theoretical expectations based on 1-D wave model (\companion). The basic cause of these asymmetries is that, as the mean field rotates away from the radial with expansion, wavevectors $\ph$ along the normal direction are on average more perpendicular to $\mB$ than those along the tangential direction. These normal-directed wavevectors then generate larger Alfv\'enic perturbations in the tangential direction, via $\ph\boldtimes\mB$. Furthermore, these tangential deflections are asymmetric as a consequence of keeping $\B^2 = {\rm const.}$ (points (iv) and (v) of \cref{sub: aw sbs}). 

Whether switchbacks within the solar wind show this asymmetric, tangentially skewed nature of deflections is still uncertain, however. In contrast to the strong skewness of deflections above, Refs.~\onlinecite{Dudok_de_Wit2020-pp},~\onlinecite{Horbury2020-bd} and \onlinecite{Laker2022-ij} report that switchbacks show preferential deflections in both the $+$T and $-$T directions (there may also be indications of this effect in the model of Ref.~\onlinecite{Schwadron2021-zt}). This may be due to a variety of factors, including different amplitudes or $\chi$ (\Cref{eq:5chidef}), or different Parker spiral angles; regardless, the general asymmetry is consistent with our arguments. The slight tangential bias of small deflections in \cref{fig:B deflections}b may be an indication of preferential deflections in both the $+$T and $-$T directions in the Alfv\'en wave model, although signs of this in a clock angle histogram as in \cref{fig:clockangle_a5} are overwhelmed by the near-isotropic nature of larger deflections. It would be interesting to see how this bias is affected by the amplitude of the fluctuations; this is left to be investigated in future work. In general, though, it is clear that the asymmetric switchback distributions cannot necessarily be associated with asymmetries of the source. 

\subsection{Compressible properties of switchbacks}\label{sub: compressible}

\begin{figure*}
\begin{center}
\includegraphics[width=\textwidth]{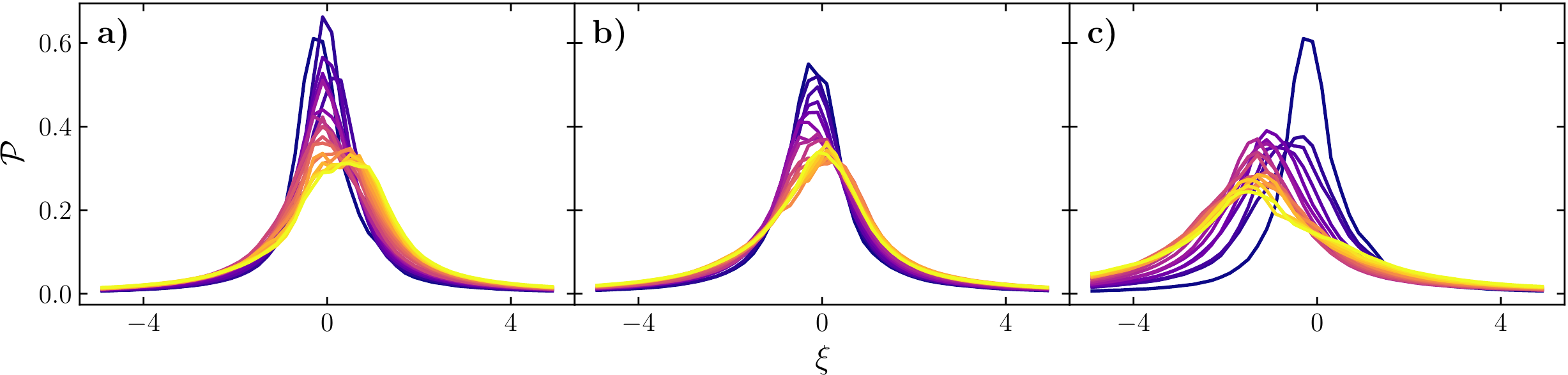}
\caption{PDFs of the polarization fraction $\xi$ (\cref{eq:polfrac}) measured within switchbacks with $z \geq 0.5$, showing its dependence on $\beta$. These PDFs are measured in the \betalow \ (a), \defaulthires \ (b), and \betahigh \ (c) simulations during their expansion from $a=1$ (purple) to $a=10$ (yellow). The evolution of $\beta$ in these simulations range from $0.16$ to $\approx 0.3$ in \betalow, $\approx 0.35$ to $\approx 0.6$ in \defaulthires, and $1$ to $\approx 2$ in \betahigh. The correlation between magnetic pressure and density fluctuations trends towards being positive for smaller values of $\beta$ (a), and towards negative for larger values (c), with a minimization for $\beta \approx 1$ (b), agreeing with the predictions made by \mallet.}
\label{fig:polfrac_evo}
\end{center}
\end{figure*}

\begin{figure}
\begin{center}
\includegraphics[width=\columnwidth]{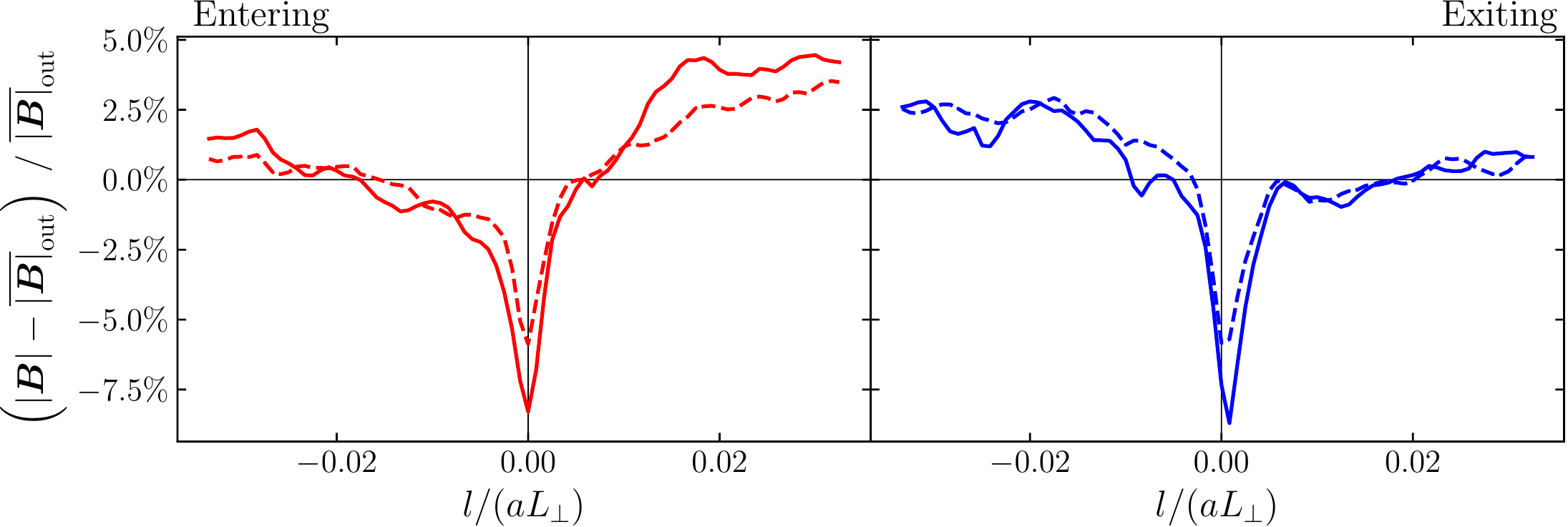}
\caption{Superposed epoch analysis of the fractional change in magnetic-field strength $|\B|$ across switchback boundaries in flybys through \defaulthires \ (solid) and \parkerhires \ (dashed) at $a=5$. Dropouts in $|\B|$ are clearly seen across boundaries, with the interior of switchbacks showing a slight increase in $|\B|$ relative to outside. The direction of travel along the flyby is from left to right.}
\label{fig:Bmag_dropouts}
\end{center}
\end{figure}

\begin{figure}
\begin{center}
\includegraphics[width=\columnwidth]{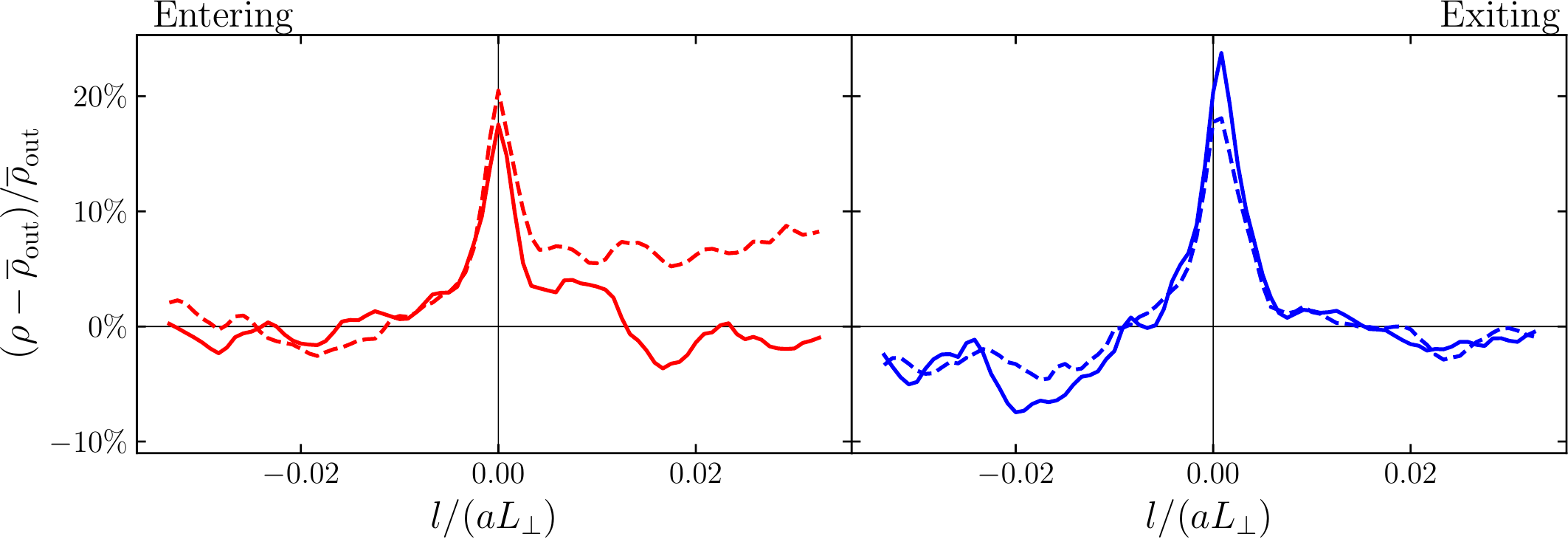}
\caption{Superposed epoch analysis of the fractional change in density $\rho$ across switchback boundaries in flybys through \defaulthires \ (solid) and \parkerhires \ (dashed) at $a=5$, showing spikes in $\rho$ relative to the average value outside switchbacks. The direction of travel is the same as in \cref{fig:Bmag_dropouts}.}
\label{fig:rho_dropouts}
\end{center}
\end{figure}

Although switchbacks primarily exhibit Alfv\'enic correlations between magnetic-field and velocity fluctuations, they also show variations in density and $\B^2$, signifying compressible behaviour. To compare with both theory and observations, we now investigate the compressible properties of switchbacks.

\subsubsection{Beta dependence of correlations between magnetic-field strength and density fluctuations} \label{subsub: beta dependence}

The statistical analysis of PSP switchback observations by Ref.~\onlinecite{Larosa2021-iy} showed that fluctuations in density and magnetic-field strength are positively correlated in some switchbacks, while being negatively correlated in others. \mallet \ argue that this may be a consequence of the expanding-Alfv\'en-wave nature of switchbacks, with their analytical theory predicting a dependence of the correlation of these fluctuations on $\beta$ and the angle of the wavevector to the background field. Although this prediction is for an Alfv\'en wave that varies only in one dimension, we can directly test whether it carries over to the complex three-dimensional case, which is clearly more directly relevant to solar-wind observations. We define the `polarization fraction' $\xi$ as 
\begin{equation}
    \xi \equiv \frac{\delta(\B^2)/\overline{\B^2}}{\delta\rho/\overline{\rho}}, \label{eq:polfrac}
\end{equation}
where $\delta f$ represents the fluctuating part of a quantity (as defined in \cref{sub: ebm theory}). In ideal MHD with no expansion, $\xi$ can be thought of as a measure of the magnetosonic waves, which have a positive (negative) correlation between magnetic and thermal pressure for the fast (slow) magnetosonic wave. When the effects of expansion are included, however, \mallet \ show that Alfv\'en waves gain a compressive component because they must continuously change shape to remain spherically polarized. \mallet \ predict that for expanding, spherically polarized Alfv\'en waves with $k_\perp \gtrsim k_\|$, $\xi > 0$ if $\beta \lesssim 1$ and $\xi < 0$ if $\beta \gtrsim 1$, with a minimization of magnetic pressure fluctuations ($\xi = 0$) at $\beta\approx 1$ for structures with $k_\perp \sim k_\|$. 

\Cref{fig:polfrac_evo} shows probability distribution functions (PDFs) of $\xi$ calculated within switchbacks with $z\geq0.5$ in the \betalow, \defaulthires, and \betahigh \ simulations. All simulations have been further run to $a=10$ in order to more clearly show the results. These simulations start out and evolve with different values of $\beta$ within the ranges considered by \mallet, with the numbers in the \betalow \ and \betahigh \ labels corresponding to the initial value of $\beta$ in these simulations which increases with expansion (as shown in \cref{fig:beta evo}a); the \defaulthires \ simulation has $\beta\approx 0.35$ initially. We see that $\xi$ trends towards positive values in \betalow \ and negative values in \betahigh, while staying closer to zero for \defaulthires. Similar trends in $\xi$ are also seen when fluctuations are measured over the entire box rather than just inside switchbacks.

A consequence of this change in polarization in the theory of \mallet \ is that fluctuations in $\B^2$ are minimized as $\beta$ passes through a critical value. We see this in the evolution of $\CB$ in \cref{fig:beta evo}b. Recall that $\CB$ is a measure of how the components of $\B$ are correlated to keep $\B^2$ constant, with smaller values of $\CB$ corresponding to smaller fluctuations in $\B^2$ within the system. The \betalow \ and \betahigh \ simulations consistently have a higher value of $\CB$ in \cref{fig:beta evo}b, with $\CB$ being minimized in \defaulthires. This provides simulation evidence for the theory of \mallet, even in complex 3-D Alfv\'enic structures.

\subsubsection{Compressible properties of switchback boundaries}\label{subsub: dropouts}

Ref.~\onlinecite{Farrell2020-md} perform a superposed epoch analysis on the properties of switchback boundaries, studying how physical quantities such as density and magnetic field components varied as PSP travelled through switchbacks. They found that the magnetic-field strength $|\B|$ remains constant throughout the switchback, except at switchback boundaries, where quick decreases in $|\B|$ they termed `dropouts' were observed. The proton density inside switchbacks was also reported to decrease relative to outside, with spikes in density often occurring at the boundaries.

To assess whether the switchbacks in the model of \emph{in-situ} Alfv\'en wave evolution have similar properties at their boundaries, we perform a similar superposed epoch analysis to Ref.~\onlinecite{Farrell2020-md} on the \defaulthires \ and \parkerhires \ simulations. Using the same flyby data as shown in \cref{fig:default_flyby}, we use a peak finding algorithm to find the locations where $|\Delta\deflectangle|$ (\cref{eq:dtheta_def}) is greater than $45^\circ$. This corresponds to the sharp rotations of the magnetic field at switchback boundaries. These boundaries are separated into switchback entries and exits (as in Ref.~\onlinecite{Farrell2020-md}), defined by $\Delta\deflectangle / |\Delta\deflectangle| = 1$ and $-1$ respectively. For each boundary, we then look at the values of the magnetic-field strength $
|\B|$ and density $\rho$ from 40 grid points on either side of the boundary, and calculate the fractional change of these quantities relative to their mean across the 40 grid points outside the switchback defined as $(|\B| - \overline{|\B|}_{\rm out})/\overline{|\B|}_{\rm out}$ and $(\rho - \overline{\rho}_{\rm out})/\overline{\rho}_{\rm out}$. A superposed epoch analysis is then performed separately for all switchback entries and exits, where the fractional changes are added together and averaged to highlight any trends across switchback boundaries.

The results of this analysis are shown at $a=5$ in \cref{fig:Bmag_dropouts} and \cref{fig:rho_dropouts}. At both the entry and exit of switchback boundaries, the magnetic-field strength $|\B|$ experiences dips relative to the mean $|\B|$ outside the switchback. Corresponding to these dropouts are spikes in density across switchback boundaries. Due to the averaging nature of the superposed epoch analysis used, common features are highlighted while random fluctuations are removed. This suggests these dropouts in $|\B|$ and spikes in density are robust features at the boundaries of switchbacks seen within our simulations.

\subsubsection{Compressible properties: Discussion}

 Switchbacks in the expanding Alfv\'en wave model exhibit compressible properties that are in reasonable agreement with both observation and theory. The evolution of the polarization fraction $\xi$ in \cref{fig:polfrac_evo} as well as the minimization of $\CB$ in \cref{fig:beta evo}b for certain values of $\beta$ shows that the $\beta$-dependent predictions of simple 1-D Alfv\'en waves in \mallet \ apply qualitatively to the complex 3-D cases seen within these expanding box simulations. This further adds support to the predictions of \mallet \ in explaining observational data (e.g. Ref.~\onlinecite{Larosa2021-iy}). 
 
 The dropouts in $|\B|$ and spikes in density across switchback boundaries are remarkably similar to those seen in switchback observations by PSP. Ref.~\onlinecite{Farrell2020-md} posit that the dropouts in $|\B|$ observed are due to a diamagnetic boundary current across that cancels the magnetic flux on either side of the switchback boundary; the dropouts we see in our simulations suggest a similar effect is likely happening in the simulations. A noticeable difference to observations is that the density within switchbacks does not decrease relative to the mean outside, as shown in \cref{fig:rho_dropouts}. However, we do not expect the density to vary exactly as in observed switchbacks, as the isothermal equation of state we use for these simulations is only an approximation to the true thermal properties of the solar wind. Future work on this subject should include a more realistic equation of state for better comparisons to data.

\section{Conclusion}\label{sec: conclusion}

In this paper, we investigate the properties of switchbacks arising from the evolution of Alfv\'en waves in the expanding solar wind outside the Alfv\'en point. High-resolution three-dimensional numerical simulations utilizing the expanding box model are initialized with an outwards-propagating collection of large-amplitude Alfv\'en waves, with this initial collection of waves exhibiting switchback-like features that evolve with expansion. The properties of these switchbacks are shown to be in good agreement with both theory and observations by Parker Solar Probe, and allow us to make further testable predictions. The key properties of switchbacks we studied in this paper can be split into two categories: asymmetries in the deflection of the magnetic field arising from the addition of a Parker spiral, and compressible properties at the boundaries of and within switchbacks.

The addition of a Parker spiral with even a small non-radial component was found to affect switchback behaviour dramatically, giving rise to asymmetrical, tangentially skewed deflections. Our companion paper \companion \ (whose results are summarized in \cref{sub: aw sbs}) investigates the behaviour of switchbacks in the Alfv\'en wave model when a Parker spiral is included, and complements the results of this paper. These properties can be summarized as follows:\vspace{0.1cm}\\
(i) Magnetic fields preferentially deflect in one direction within switchbacks in a Parker spiral; switchbacks with rotations more than $90^\circ$ from the mean field exhibit this most strongly (\cref{subsub: clock angle}; point (v) of \cref{sub: aw sbs}). These deflections are `tangentially skewed': they point in the tangential direction towards the radial component of the background magnetic field. In contrast, switchbacks in a radial background field are necessarily symmetric. Observations of switchbacks seem to show a preference for deflections in the tangential direction \cite{Dudok_de_Wit2020-pp, Horbury2020-bd,Laker2022-ij}, although whether these are asymmetric is uncertain; this can be tested with further switchback observations by PSP.\vspace{0.1cm}\\
(ii) In the distribution of magnetic field deflections with a large Parker angle, the most probable direction is aligned further towards the tangential direction than the Parker spiral (or mean field) direction (\cref{subsub: B deflections}; point (vii) of \cref{sub: aw sbs}). If this is true within solar-wind data from PSP, the most common field direction may differ significantly from the Parker spiral direction, which is the direction that fluctuations propagate.\vspace{0.1cm}\\
(iii) Switchbacks within a Parker spiral tend to exhibit sharper rotations than in the case of a radial field (\cref{fig:rotation_sharpness}; point (vi) of \cref{sub: aw sbs}).\vspace{0.1cm}\\
(iv) The addition of a Parker spiral appears to enhance the growth of switchbacks with expansion (\cref{sub: sbfrac evo}; point (iii) of \cref{sub: aw sbs}); however, the effects of strong turbulent decay (summarized below) can stop this growth.

The compressible properties of switchbacks in this model can be summarized as follows:\vspace{0.1cm}\\
(i) Correlations between magnetic-field-strength and density fluctuations within switchbacks follow the $\beta$-dependent predictions of Ref.~\onlinecite{Mallet2021-fh} (\cref{subsub: beta dependence}). This shows that the properties of their model of Alfv\'enic switchbacks also carry over to the complex, 3-D simulations in this paper, which are more representative of the solar wind. This lends further support to the predictions of Ref.~\onlinecite{Mallet2021-fh} in explaining observational data.\vspace{0.1cm}\\
(ii) The near constant magnetic-field strength within simulations exhibits sharp `dropouts' at switchback boundaries, as well as spikes in density (\cref{subsub: dropouts}). These are akin to those reported in switchbacks observed by PSP \cite{Farrell2020-md}, and are likely due to diamagnetic currents.\vspace{0.1cm}\\

The use of the expanding box model -- with its assumption of constant solar-wind velocity -- limits the applicability of these results to outside the Alfv\'en point, where turbulent behaviour can stagnate the growth of the normalized amplitude of fluctuations and stop the growth of switchbacks. The normalized amplitude can grow inside the Alfv\'en point regardless of turbulent decay, however, allowing us to imagine the simulations start out with large-amplitude Alfv\'en waves propagating outwards from the Alfv\'en point. Further investigations into the formation of switchbacks via Alfv\'en waves need to use a model that can capture the evolution of waves inside the Alfv\'en point, such as the accelerating expanding box of Ref.~\onlinecite{Tenerani2017} or flux-tube simulations like those of Ref.~\onlinecite{Shoda2021-ld}.

We stress that the properties of switchbacks in this paper arise naturally from the evolution of Alfv\'enic structures \emph{in-situ}: our simulations are initialized with a random collection of large-amplitude, outwards-propagating Alfv\'en waves with no assumptions of influence from solar-surface processes. Because of this, the results of this paper can be tested against observations to help differentiate between the influences of \emph{in-situ} and \emph{ex-situ} processes on the properties of switchbacks within the solar wind.

\begin{acknowledgments}
The authors thank R.~Laker and T.~Horbury for interesting discussions about observational data over the course of this work. Support for Z.J.~was provided by a postgraduate publishing bursary from the University of Otago. Support for J.S.~was provided by Rutherford Discovery Fellowship RDF-U001804, which is managed through the Royal Society Te Ap\=arangi, and R.M.~was supported by Marsden fund grant MFP\_U0020 and Rutherford Discovery Fellowship RDF-U001804. High-performance computing resources were provided by the New Zealand eScience Infrastructure (NeSI) under project grant uoo02637.
\end{acknowledgments}

\section*{Author Declarations}

The authors have no conflicts to disclose.

\section*{Data Availability Statement}

The data that support the findings of this study are available from the corresponding author upon reasonable request.

\appendix*

\section{HLLD Riemann Solver Implementation for the EBM} \label{app:riemman solver}

\subsection{Equations}

We use a modified version of the HLLD Riemann solver of Ref.~\onlinecite{Mignone2007-lb} to solve \cref{eq:ebm_all} within the expanding box frame. We use the variables \cite{Hellinger2005-ae, Bott2021-vp}
\begin{equation}
\rho' = \lambda\rho, \quad \vel' = \BLambda^{-1}\bolddot\vel, \quad \B' = \lambda\BLambda^{-1}\bolddot\B, \quad \tildenabla = \BLambda^{-1}\bolddot\nabla',\label{eq:ebm_varchange}
\end{equation}
where $\BLambda=\text{diag}(1, a, a)$ is a matrix representing expansion along the $y$ and $z$ directions, $\lambda \equiv \text{det }\BLambda = a^2$, and $\nabla'$ is the expansion-free gradient. This change of variables removes the expansion source terms in \cref{eq:ebm_density,eq:ebm_induction}, bringing them into an ideal MHD-like form
\begin{equation}
    \parfrac{\rho'}{t} + \nabla' \bolddot (\rho'\vel') = 0\label{eq:Aebm_density_modified}
\end{equation}
and
\begin{equation}
    \parfrac{\B'}{t} -\nabla'\boldtimes(\vel'\boldtimes\B') = 0.\label{eq:Aebm_induction_modified}
\end{equation}
All the effects of expansion are moved into the momentum equation, which becomes
\begin{equation}
    \parfrac{(\rho'\vel')}{t}+ \nabla'\bolddot\bsf{G} = -2\dot{\BLambda}\BLambda^{-1}\bolddot(\rho'\vel'),\label{eq:Aebm_momentum_modified}
\end{equation}
where the stress tensor
\begin{equation}
    \bsf{G} = \rho'\vel'\vel' + \left(c^2_s\rho' + \frac{1}{\lambda}\frac{(\BLambda\bolddot\B')^2}{8\pi}\right)(\BLambda^{-1})^2 - \frac{1}{\lambda}\frac{\B'\B'}{4\pi}.
\end{equation}
Ref.~\onlinecite{Johnston2022-ig} gives more details on the derivation of these equations and their stability for large expansion factors.

\subsection{Modifying fluxes within the HLLD solver}

The HLLD isothermal MHD Riemann solver developed by Ref.~\onlinecite{Mignone2007-lb} used in \athenapp \ calculates the fluxes through faces normal to the $x$-, $y$- and $z$-directions one at a time. This gives rise to a one-dimensional conservative equation ${\partial\bm{U}/\partial t + \partial\bm{F}/\partial x = \bm{0}}$, where $\bm{U}$ is a vector of relevant quantities and $\bm{F}$ is a vector of the fluxes of these quantities in the $x$-direction. For the EBM with the change of variables in \cref{eq:ebm_varchange}, this gives
\begin{equation}
    \bm{U} = \begin{pmatrix}
            \rho' \\
            \rho' u'_x \\
            \rho' u'_y\\
            \rho' u'_z \\
            B'_x \\
            B'_y \\
            B'_z 
         \end{pmatrix}, \quad 
    \bm{F} = \begin{pmatrix}
            \rho' u'_x\\
            \rho' u'^{2}_{x} + p'_{{\rm tot}, i} - B^{\prime 2}_x / a^2  \\
            \rho' u'_x u'_y - B'_x B'_y / a^2\\
            \rho' u'_x u'_z - B'_x B'_z / a^2 \\
            0 \\
            B'_y u'_x - B'_x u'_y \\
            B'_z u'_x - B'_x u'_z 
         \end{pmatrix}, \label{eq:Auandf_mod}
\end{equation}
where $\B'/\sqrt{4\pi} \to \B'$ is used for simplicity of notation. The quantity ${p'_{{\rm tot}, i}\equiv \alpha^{-1}_i (a^2c_s^2\rho' + (\BLambda\bolddot\B')^2/2)}$ is the modified total pressure where $i=x,y,z$ represents the direction the Riemann solver is calculating the fluxes in, with $\alpha_x = a^2$ and $\alpha_y=\alpha_z=a^4$. The source terms in the momentum equation (right-hand side of \cref{eq:Aebm_momentum_modified}) are added on after the fluxes have been calculated.

The speeds of the fast magnetosonic and Alfvén waves, which are important in determining the fluxes through the cell boundary as well as calculating the CFL condition required for stability, are also modified when using this form of the equations. The Alfvén wave speed within the solver, $\va= \B / \sqrt{\rho}$, is simply multiplied by a factor of $a^{-1}$ as in the scalings given by the EBM (\cref{sub: ebm theory}). In contrast, the expression for the fast magnetosonic speed $c_f$ in this new implementation is split into three cases depending on the direction of the solver, and is given by
\begin{equation}
    c_{f,i} = \sqrt{\frac{p_{B,i} + \gamma_i c_s^2\rho' + \sqrt{(p_{B,i} - \gamma_i c_s^2\rho')^2 - 4\gamma^2_i B'^{2}_i c_s^2\rho'}}{2\rho'}}. \label{eq:Afastmagspeed}
\end{equation}
Here, $\gamma_x = 1$, $\gamma_y = \gamma_z = a^{-2}$, and $p_{B, i} \equiv \gamma_i(a^{-2} B'^{2}_x + B'^{2}_y + B'^{2}_z)$ is the solver-direction-dependent magnetic pressure within the fast wave speed. The sound speed at the current point of expansion is given by $c_s(a) = c_{s0}a^{-2/3}$, with $c_{s0}$ the initial sound speed.

\bibliographystyle{aipnum4-1}
\bibliography{main}


\end{document}